\documentclass[twocolumn,showpacs,preprintnumbers,amsmath,amssymb,prl]{revtex4}

\usepackage{graphicx}
\usepackage{bm}

\begin{document}

\title{Complex Phase Behavior of The System of Particles
with Smooth Potential with Repulsive Shoulder and Attractive Well}

\author{Yu. D. Fomin}
\affiliation{Institute for High Pressure Physics, Russian Academy
of Sciences, Troitsk 142190, Moscow Region, Russia}

\author{V. N. Ryzhov}
\affiliation{Institute for High Pressure Physics, Russian Academy
of Sciences, Troitsk 142190, Moscow Region, Russia}

\author{E. N. Tsiok}
\affiliation{Institute for High Pressure Physics, Russian Academy
of Sciences, Troitsk 142190, Moscow Region, Russia}

\date{\today}

\begin{abstract}
We report a detailed simulation study of the phase behavior of
core softened system with attractive well. Different repulsive
shoulder widthes and attractive well depthes are considered which
allows to monitor the influence of repulsive and attractive forces
on the phase diagram of the system. Thermodynamic anomalies in the
systems are also studied. It is shown that the diffusion anomaly
is stabilized by small attraction.
\end{abstract}

 \maketitle

\section{I. Introduction}

In 1970-th Hemmer and Stell introduced a model system which can
effectively approximate real substances \cite{hemmer,stell}. The
model they introduced belongs to the class of potentials which
nowadays is called core-softened potentials, i.e. the system with
softening of the repulsive part of interparticle interaction.
Hemmer and Stell showed that such systems can demonstrate some
unusual phase transitions - transition between two crystal phases
of the same symmetry \cite{hemmer,stell} which ends in a critical
point similar to the gas-liquid one, and liquid - liquid phase
transition. In Ref. \cite{cecs} it was proposed that the high
density transition with critical point qualitatively represents
the isostructural transition in Ce and Cs.

Great interest to this kind of systems appeared after Hemmer and
Stell publications. Many researchers studied this model or
proposed other core-softened model potentials.

Young and Alder carried out simulation of collapsing spheres (CS)
system which is defined as $U(r)=\infty$ if $r<\sigma$,
$U(r)=\varepsilon$ if $\sigma<r<\sigma_1$ and $U(r)=0$ otherwise
\cite{ya11,ya12}. Because of the computer power restrictions they
managed to estimate the melting line of the system only in
two-dimensional case. According to their's results the melting
line demonstrates a maximum. They pointed out the similarity of
the phase diagrams of the collapsing spheres system and
experimental phase diagram of Ce and Cs.

However, Young and Alder did not consider solid-solid phase
transformation of collapsing spheres. This was done by Bolhuis and
Frenkel who studied the CS system by Monte-Carlo simulation
\cite{bf}. They considered isostructural FCC-FCC transition in the
collapsing spheres and square well systems. As it was shown in the
cited paper, the isostructural transition takes place in both
systems for sufficiently small soft core or square well widths. If
the width of the soft core (well) increases the solid-solid
transition disappears.

Velasco et al. studied the collapsing spheres system in the frame
of thermodynamic perturbation theory \cite{cpbiri}. Their's
results are in good agrement with the Bolhuis and Frenkel
publication \cite{bf}. They considered four values of the step
width: $\sigma_1=1.03;1.08;1.1$ and $1.16$. According to their's
findings the first three systems demonstrate the isostructural
solid-solid transformation with critical point while the last
system does not have this transition. At the same time an
FCC-BCC-FCC sequence occurs in the system with the largest step
size.

The qualitative phase behavior of collapsing spheres was studied
by Stishov \cite{st}. This article makes a qualitative estimation
of the collapsing spheres phase diagram basing on the high- and
low- temperature limits. It was shown that depending on the
$\sigma_1 / \sigma$ ratio the phase diagram can change
dramatically. This qualitative consideration is in good agreement
with Monte-Carlo simulation \cite{bf} and perturbation theory
results \cite{cpbiri} mentioned above. In Refs.
\cite{RS2002,RS2003} it was discussed the possibility of the
liquid-liquid phase transition in CS system.

One can use the geometrical reasons to identify the possible
structures of the core softened systems. This method was applied
by Jagla to a set of systems \cite{jaglajcp}. In this paper a
parameter dependent potential was considered. Depending on the
value of the tuning parameter it can be continuously changed from
a hard core plus linear ramp shape to hard core plus repulsive
shoulder one. The minimal energy configurations corresponding to
different pressures for the different potentials were monitored.
It was established that depending on the pressure the number of
overlaps of the soft core changes producing a large number of
possible geometrical configurations (Fig. 3 in \cite{jaglajcp}).
Basing on this argument one can expect a rich variety of
structures in the core softened systems both in crystalline and
disordered regions.

Although the main focus of Ref. \cite{jaglajcp} was on
two-dimensional systems, the results were also generalized for
three-dimensional case (Fig. 6 in \cite{jaglajcp}). These results
clearly state that the crystalline region of the linear ramp
system demonstrates many phases with different symmetry. This fact
identifies that core-softening is a very strong mechanism
affecting the phase behavior of the system.

The minimum energy configurations in the two-dimensional
collapsing spheres system was revised later in Refs.
\cite{MalescioPellicane1,MalescioPellicane2}. In this paper the MC
simulation of the system was also carried out and complex phase
behavior in liquid phase was observed. Depending on the density,
the system can form different cluster phases. In particular, long
stripe clusters were observed at some densities and temperatures.

The same system of two-dimensional collapsing spheres was studied
further in Ref. \cite{NorizoeKawakatsu}. The stripe phases for a
set of the step widths were observed. The authors pointed out that
the stripe phase takes place only for some particular values of
the step width. Although the paper was mainly focused on the
two-dimensional system the authors claim that the conclusions are
applicable for the three dimensional case too.

Another soft core softened model in two dimensions was considered
by Camp \cite{Camp,Campdyn}. Phase diagram obtained from NPT-MC
was reported. This phase diagram consists from low- and high
density solid phases, normal liquid and cluster liquid. The
cluster liquid represents the stripe phase.

Interestingly, the stripe phases were observed experimentally in
the system of colloids in magnetic field \cite{csexp}. The
configurations observed in this system closely resemble the ones
obtained by Camp \cite{Camp}. This justifies that the
core-softened models can be successfully applied to the real
experimental systems.

A strong method of the ground state phase calculation was
developed in Refs. \cite{cst0,cst01}. This method allows to find
the minimal energy configurations by applying generic algorithms
carrying out the minimization. The collapsing sphere system with
different step widths was considered in Refs. \cite{cst0,cst01}.
Three values of the step were considered: $\sigma_1=1.5$,
$\sigma_1=4.5$ and $\sigma_1=10.0$. It was shown that the
complexity of the system increases dramatically with increasing
the step size. For example, for the system with $\sigma_1=1.5$ the
number of phases reported in \cite{cst0} is $6$, for
$\sigma_1=4.5$ - $33$ and for $\sigma_1=10.0$ - $47$.

In our recent paper a smoothed repulsive shoulder system (RSS) was
introduced \cite{wejcp}. This model is defined by the
interparticle potential
$$U(r)=\left(\frac{\sigma}{r}\right)^{14}+\frac{1}{2}\varepsilon
(1-\tanh(k_0[r-\sigma_1])),$$ where $k_0=10.0$ and the step width
$\sigma_1$ was varied. The dependence of the phase diagram on
$\sigma_1$ was studied. The following values of $\sigma_1$ were
considered: $1.15$, $1.35$, $1.55$ (Ref. \cite{wejcp}) and $1.80$
(Ref. \cite{wepre}). In agreement with previous discussion, the
behavior of the system was found to be very complex.

If the step width is relatively small then the crystalline region
consists of low- and high density FCC phases separated by a BCC
region. However, already at $\sigma_1=1.35$ the complexity of the
phase diagram greatly increases. It is important to note that
several crystalline structures appear in the system including not
close packed structures which can not be observed in LJ-like
systems. Comparing the phase diagrams for $\sigma_1=1.35$, $1.55$
and $1.8$, one can see that the complexity of the phase diagram
even increases with increasing the step width.

Later the ground state of the RSS with $\sigma_1=1.35$ was studied
in the work \cite{fominpot}. The set of the phases reported in
this work is different from the one in \cite{wejcp}. It can be
related to the different choice of the phases the authors of these
papers considered to be potentially stable.

One can see that the behavior of the purely repulsive
core-softened models is rather complex. However, the potentials of
interaction between real particles have both repulsive and
attractive parts. To take it into account one should add an
attractive well to the repulsive core-softened model. The
potentials of this type were reported as models of water, liquid
metals, colloidal particles or polymer colloid mixtures and so on.
This makes them important for the soft matter science and attracts
much attention to such systems.

The simplest core-softened system with attraction is the
collapsing spheres system with additional attractive well (CSAW):

\begin{equation}
\Phi (r)=\left\{
\begin{array}{lll}
\infty , & r\leq \sigma_0 \\
 \varepsilon_1 , & \sigma_0 <r\leq \sigma_1  \\
 -\varepsilon_2 , & \sigma_1 <r\leq \sigma_2  \\
0, & r> \sigma_2 
\end{array}%
\right. . \label{1}
\end{equation}

Liquid-liquid phase transition (LLPT) was found in such systems.
LLPT in CSAW systems was intensively studied by molecular
dynamics, integral equations theory and perturbation theory
\cite{ll1,ll2,ll3,ll4,ll5,FRT2006}. It was found that LLPT takes
place only at some specific values of parameters $\varepsilon_2 /
\varepsilon_1$, $\sigma_1 / \sigma_0$ and $\sigma_2 / \sigma_0$.
However, as far as we know, all papers on this topic suggest that
LLPT is metastable with respect to crystallization. This makes
extremely important to know also the melting line of the system.
However, just a few publications consider the crystalline region
of the phase diagram \cite{stwell1,stwell2}.

In the paper \cite{buld66} the LLPT in CSAW was studied. The
authors also investigated possible crystal structures in the
system. They came to the conclusion that only one crystal phase is
possible in the range of densities they considered. Taking into
account huge variety of phases in the purely repulsive systems
this conclusion looks rather surprising and makes it necessary to
carry out further investigations of the influence of repulsive and
attractive forces on the phase diagram of a model system.

A detailed study of the phase diagrams of a family of
core-softened systems with attractive well was carried out by
Quigley and Probert \cite{stwell1,stwell2}. The potential used in
their's works consisted of Lennard-Jones part plus a gaussian
well. Choosing different parameters they changed the depth of the
well and monitored the changes in the phase diagram. It was shown
that FCC is not the only stable crystal structure for this kind of
potentials. Simple hexagonal structure was also reported
\cite{stwell1,stwell2}.

The goal of this paper is to provide a detailed study of the
influence of attractive and repulsive forces on the phase diagram
of a core-softened system. For doing this we carry out the
calculations of the phase diagrams of a set of systems which have
the same functional form of the interatomic potential but
parameters of this potential are different. In the present article
we focus on the melting line of the system and the transformations
between different crystal phases. We also consider the influence
of attraction on the thermodynamic anomalies in the core softened
system.

\section{II. System and Methods}

In the present study a system of particles interacting via the
potential with "hard" core, repulsive shoulder and attractive well
is investigated. This potential represents a generalization of our
previous RSS model \cite{wejcp,wepre} and we call it repulsive
shoulder system with attractive well (RSS-AW) potential.

The general form of the potential is written as

\begin{equation}
  U(r)=\varepsilon
  (\frac{\sigma}{r})^{14}+\lambda_0- \lambda_1\tanh(k_1\{r-\sigma_1\})+\lambda_2
  \tanh(k_2\{r-\sigma_2\}).
\end{equation}

\begin{figure}
\includegraphics[width=8cm, height=8cm]{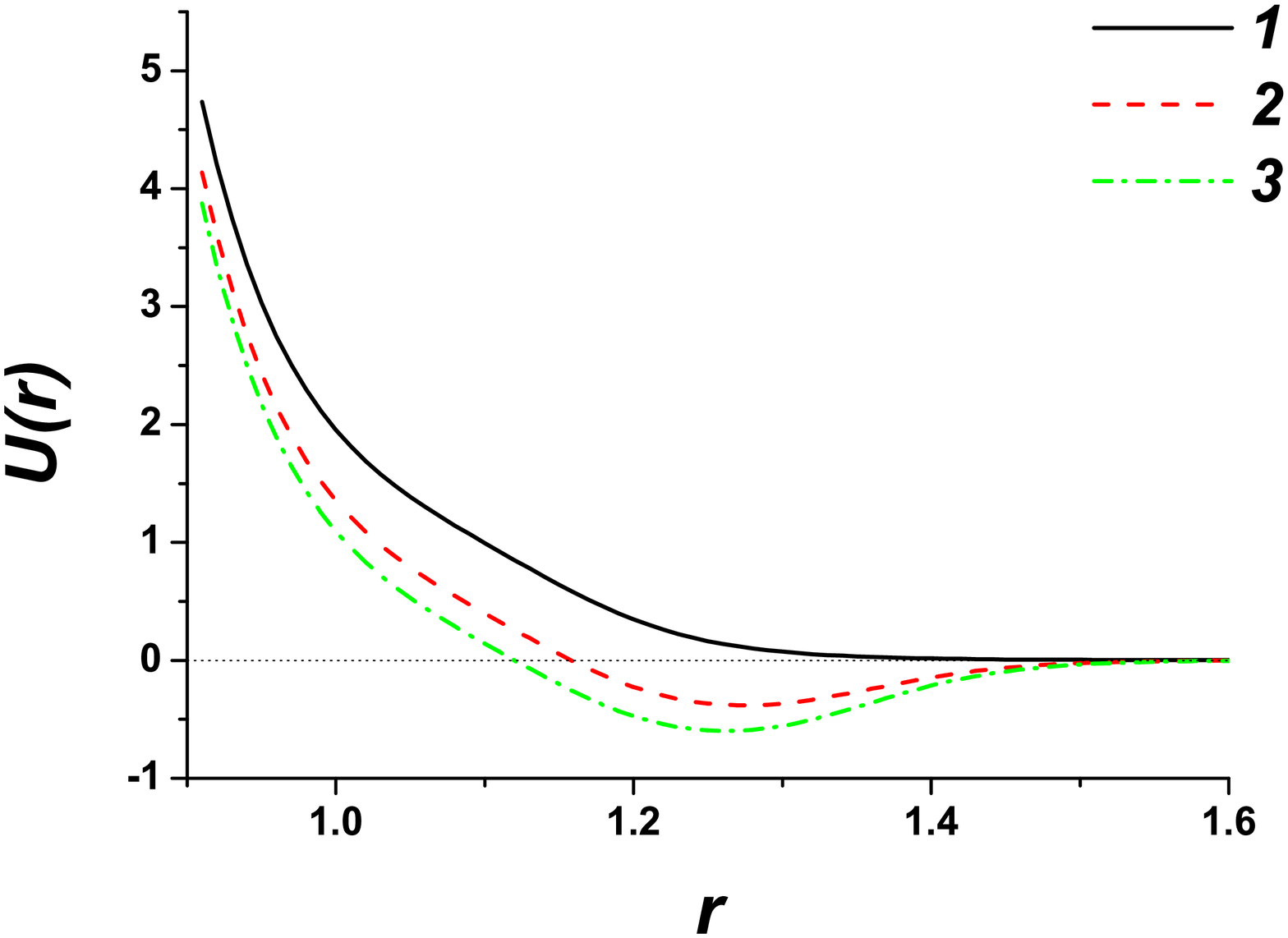}%

\caption{\label{fig:fig1} (Color online) Family of the potentials
with $\sigma_1=1.15$ and different attractive wells. The curves
are numerated in accordance with Table 1.}
\end{figure}

\begin{figure}
\includegraphics[width=8cm, height=8cm]{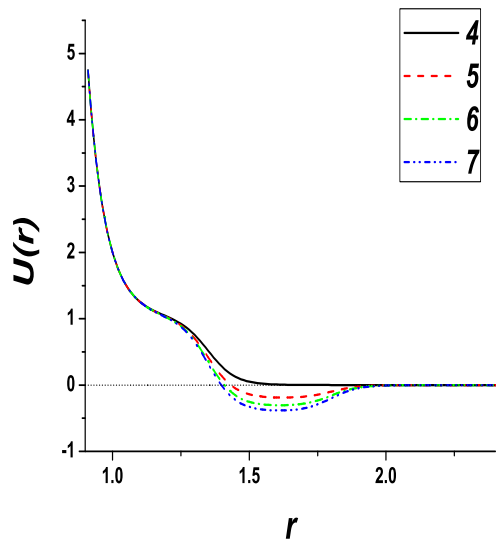}%

\caption{\label{fig:fig2} (Color online) Family of the potentials
with $\sigma_1=1.35$ and different attractive wells. The curves
are numerated in accordance with Table 1.}
\end{figure}

Two sets of systems are considered: the ones with the step width
equal to $\sigma_1=1.15$ and the ones with the step
$\sigma_1=1.35$. The parameters $k_1=k_2=10.0$ are fixed while
parameters $\sigma_1$, $\sigma_2$, $\lambda_0$, $\lambda_1$ and
$\lambda_2$ are varied to get the different potential shape. Seven
sets of parameter are considered. They are summarized in Table 1.
Figs. ~\ref{fig:fig1} and ~\ref{fig:fig2} show the potentials for
the step width $\sigma_1=1.15$ and $\sigma_1=1.35$ respectively.
The parameters are chosen in such a way that the depth of
attractive well becomes larger (see Table 1 and Figs.
~\ref{fig:fig1} and ~\ref{fig:fig2}). Below we denote the systems
with different parameters as system 1, system 2 and so on in
accordance with the Table 1.

\begin{table}
\begin{tabular}{|c|c|c|c|c|c|c|}
  \hline
  number & $\sigma_1$ & $\sigma_2$& $\lambda_0$  & $\lambda_1$ & $\lambda_2$ & well depth \\
  \hline
  1 & 1.15 & 0 & 0.5 & 0.50 & 0 & 0\\
  2 & 1.15 & 1.35 & 0.2 & 0.5 & 0.3 & 0.4\\
  3 & 1.15 & 1.35 & 0.07 & 0.5 & 0.43 & 0.60\\
  4 & 1.35 & 0 & 0.5 & 0.5 & 0 & 0\\
  5 & 1.35 & 1.80 & 0.5 & 0.60 & 0.10 & 0.20\\
  6 & 1.35 & 1.80 &0.5 & 0.66 & 0.16 & 0.30\\
  7 & 1.35 & 1.80 & 0.5 & 0.7 & 0.20 & 0.4\\
  \hline

\end{tabular}

\caption{The potential parameters used in simulations (Eq. (2)).}

\end{table}

In our previous publications \cite{wejcp,wepre} we discussed the
phase diagrams of several purely repulsive systems, i.e. the
systems with zero well depth. The complexity of these phase
diagrams was shown. Here we extend this study to the systems with
attractive well and study the evolution of the system behavior
with increasing attraction.

In order to get a hint about the possible crystal phases in the
systems we measure the energies of ideal crystal lattices which
correspond to zero temperature free energies. After that we
simulate the most preferable structures at finite temperature and
monitor the limits of theirs stability by looking at the radial
distribution functions. This procedure allows to find the
approximate regions of the thermodynamic parameters for free
energy computations which are necessary for the exact
determination of the transition lines.

We simulate the system in $NVT$ ensemble using Monte-Carlo method.
The number of particles in the liquid or gas state simulation was
set to $500$ or $1000$ and for crystal phases it varied between
$500$ and $1000$ depending on the structure. The system was
equilibrated for $10^6$ MC step and the data were collected during
$10^5$ MC steps.

In order to find the transition points we carry out the free
energy calculations for different phases and construct a common
tangent to them. In our previous article where we considered the
purely repulsive potentials we computed the free energy of the
liquid by integrating the equation of state along an isotherm
\cite{bib}:
$\frac{F(\rho)-F_{id}(\rho)}{Nk_BT}=\frac{1}{k_BT}\int_{0}^{\rho}\frac{P(\rho')-\rho'
k_BT}{\rho'^2}d\rho'$. In the case of potentials which contain an
attractive part the situation is more complicated because of the
possible gas - liquid transition. In order to avoid the
difficulties connected to this transition we carry out calculation
of free energies at high temperature above the gas - liquid
critical point and then calculate the free energies by integrating
the internal energies along an isochor \cite{bib}:
$\frac{F(T_2)-F(T_1)}{k_BT}=\int_{T_1}^{T_2}U(T,N,V)d(\frac{1}{T})$.

Free energies of different crystal phases were determined by the
method of coupling to the Einstein crystal \cite{bib}.

To improve the statistics (and to check for internal consistency)
the free energy of the solid was computed at many dozens of
different state points and fitted to multinomial function. The
fitting function we used is $a_{p,q}T^pV^q$, where $T$ and
$V=1/\rho$ are the temperature and specific volume and powers $p$
and $q$ are related through $p+q \leq N$. The value $N$ we used
for the most of calculations is $5$.

In this paper we use the dimensionless quantities: $\tilde{{\bf
r}}={\bf r}/ \sigma$, $\tilde{P}=P \sigma
^{3}/\varepsilon ,$ $\tilde{V}=V/N \sigma^{3}=1/\tilde{\rho},$ $\tilde{T}%
=k_{B}T/\varepsilon $. Since we use only these reduced units we
omit the tilde marks.

\section{III. Phase Diagrams}

\subsection{$\sigma_1=1.15$ (potentials 1 - 3)}

\begin{figure}
\includegraphics[width=8cm, height=8cm]{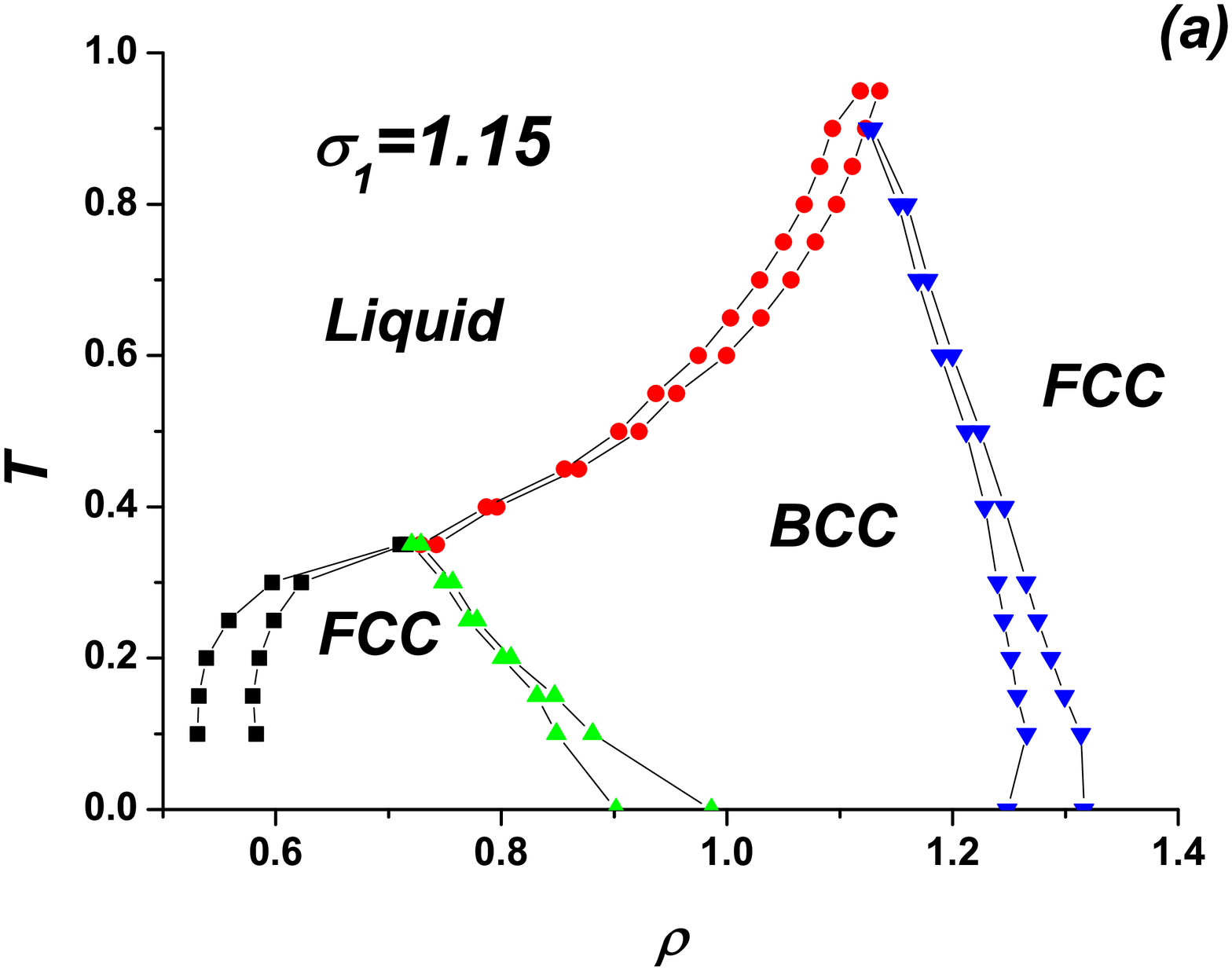}%

\includegraphics[width=8cm, height=8cm]{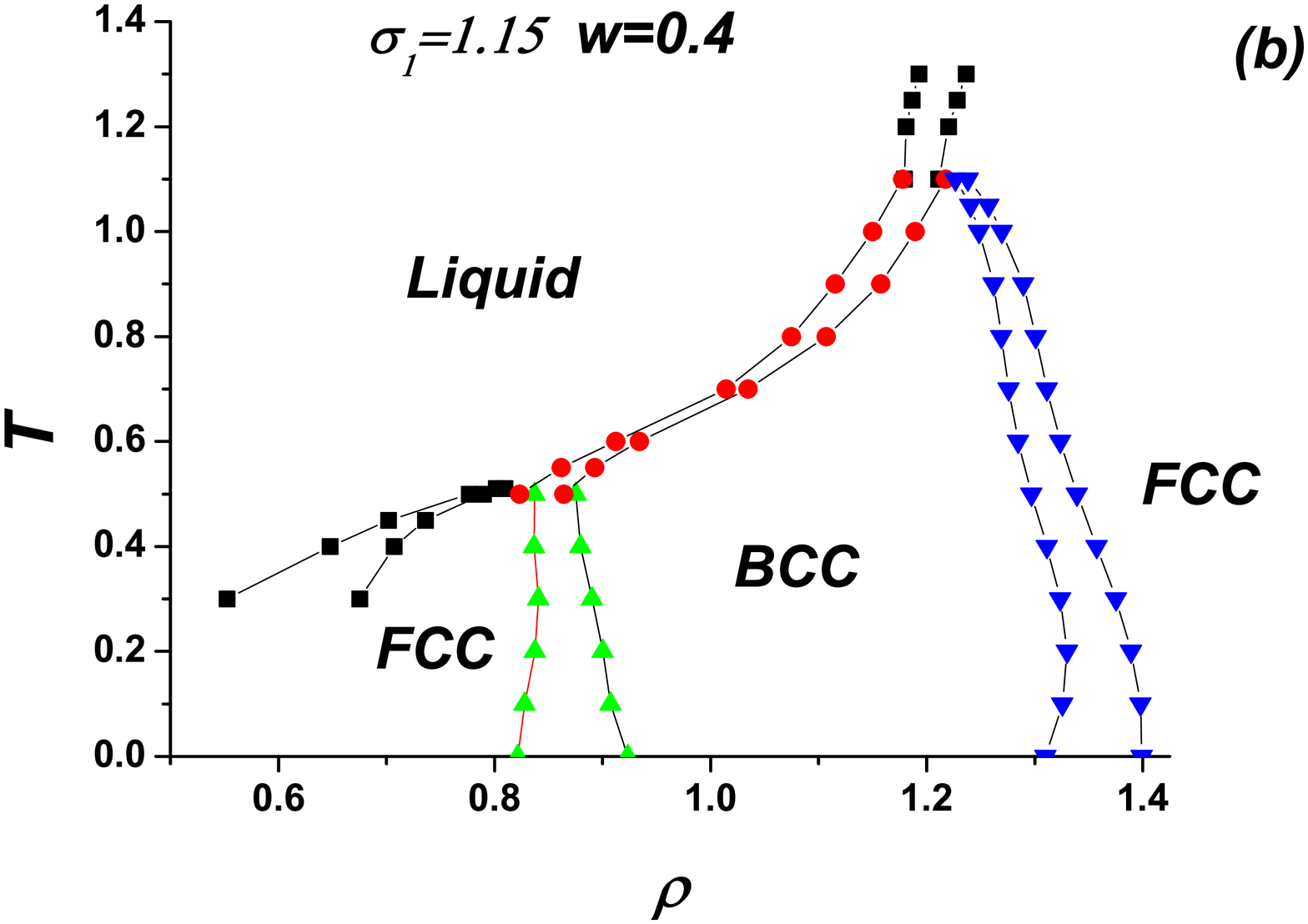}%

\includegraphics[width=8cm, height=8cm]{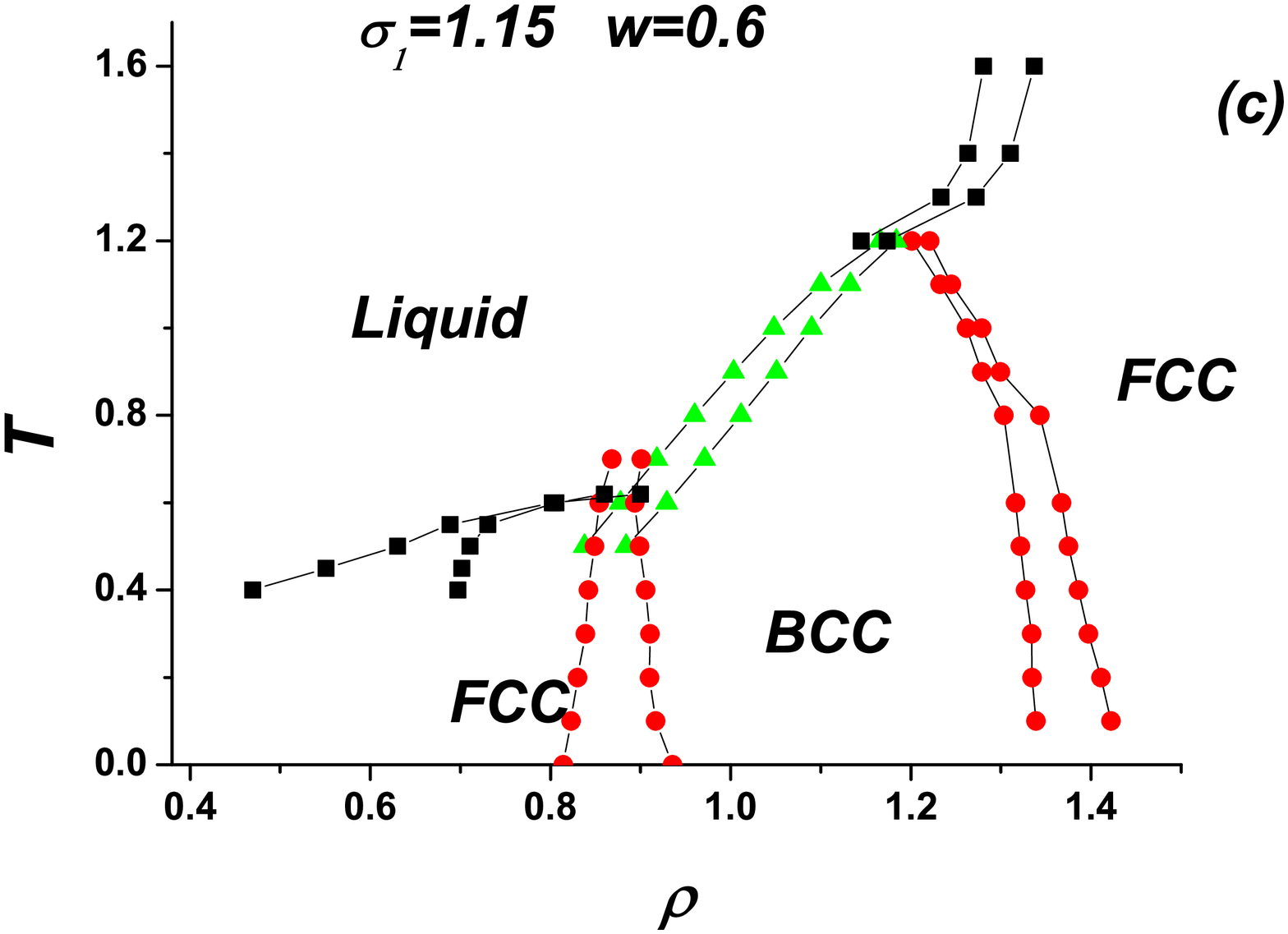}%

\caption{\label{fig:fig3} (Color online) Phase diagrams of the
systems with $\sigma_1=1.15$ (system 1 - 3) in $\rho - T$
coordinates. a - system 1, b- system 2, c - system 3.}
\end{figure}

\begin{figure}
\includegraphics[width=8cm, height=8cm]{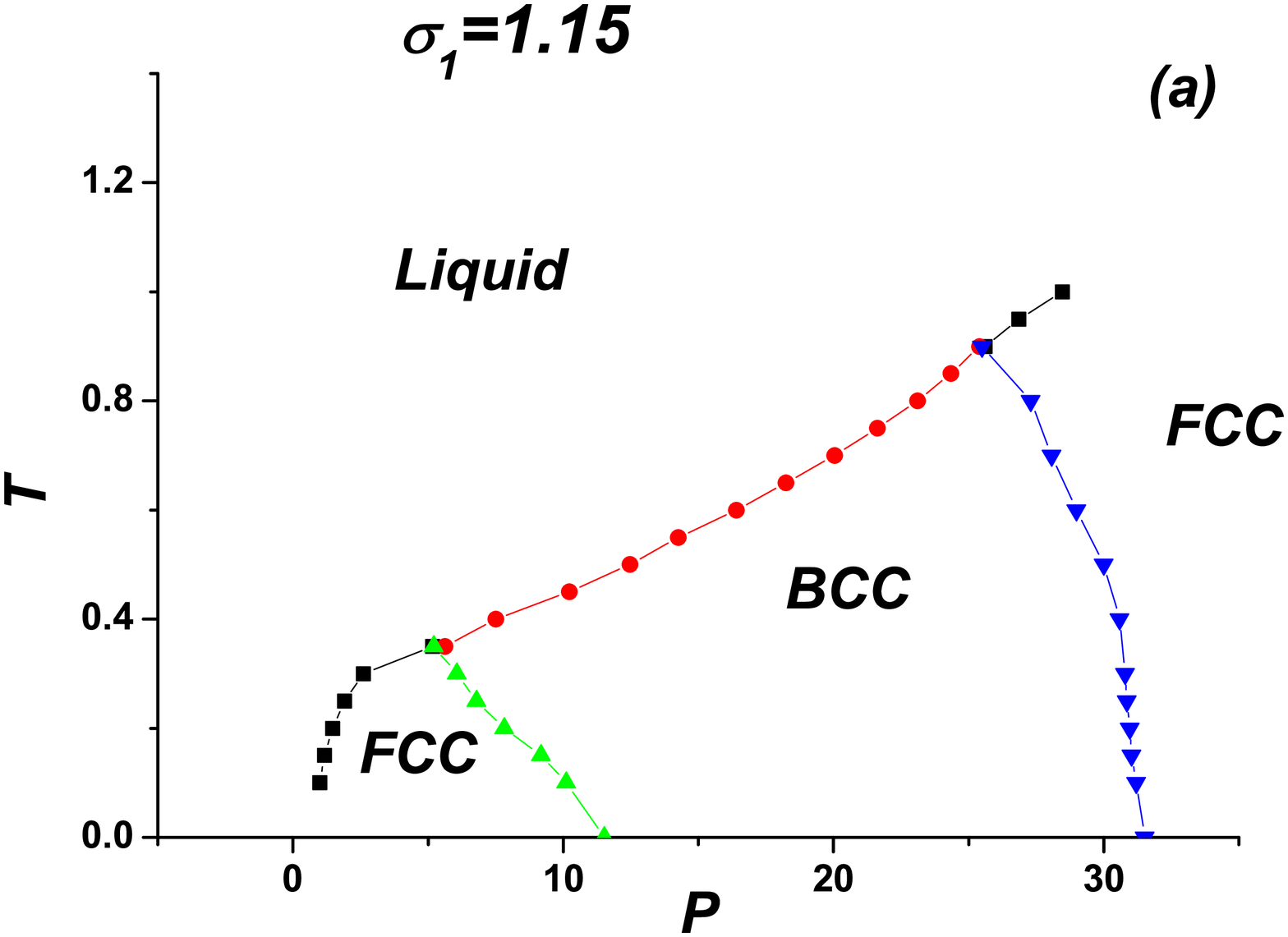}%

\includegraphics[width=8cm, height=8cm]{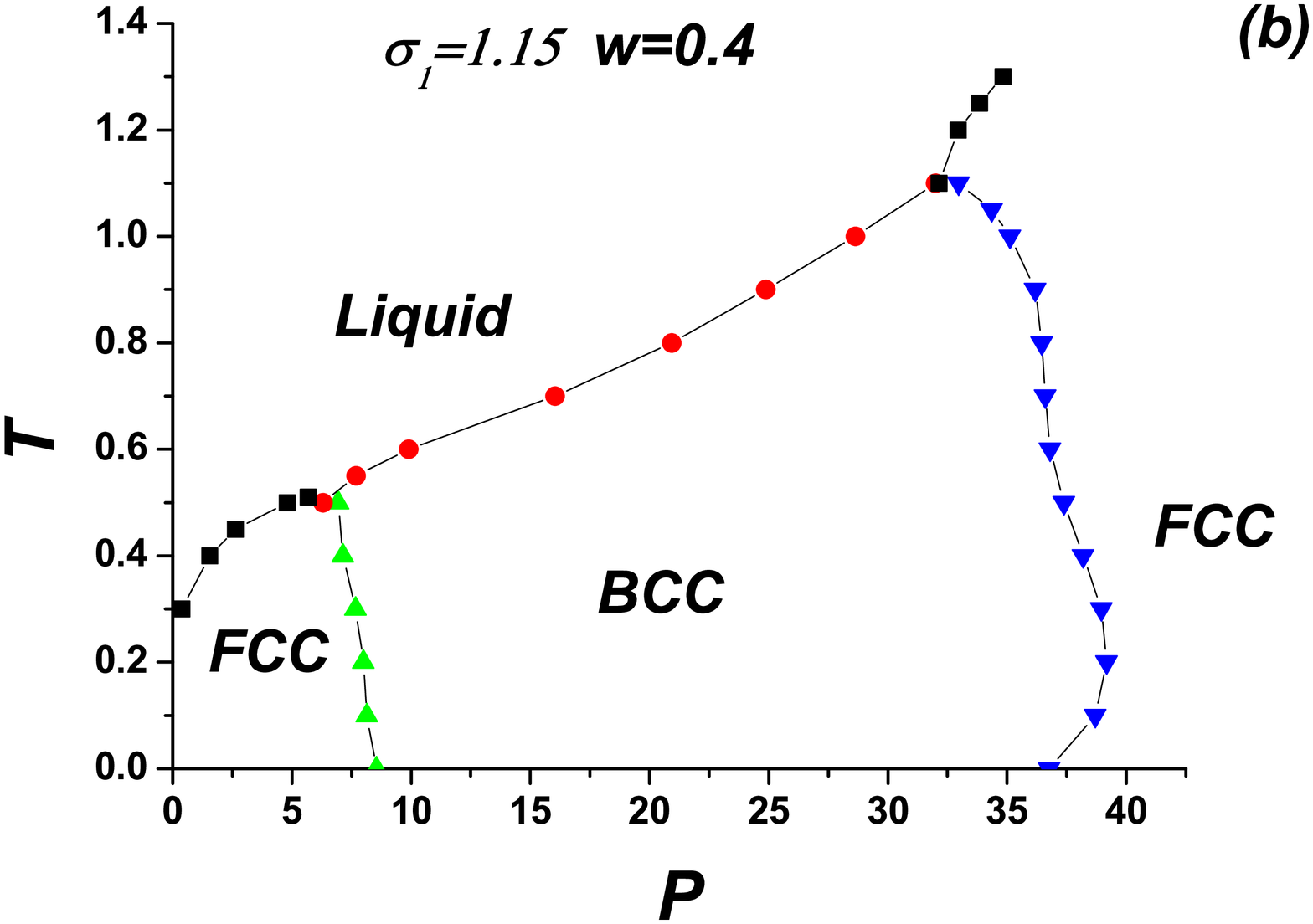}%

\includegraphics[width=8cm, height=8cm]{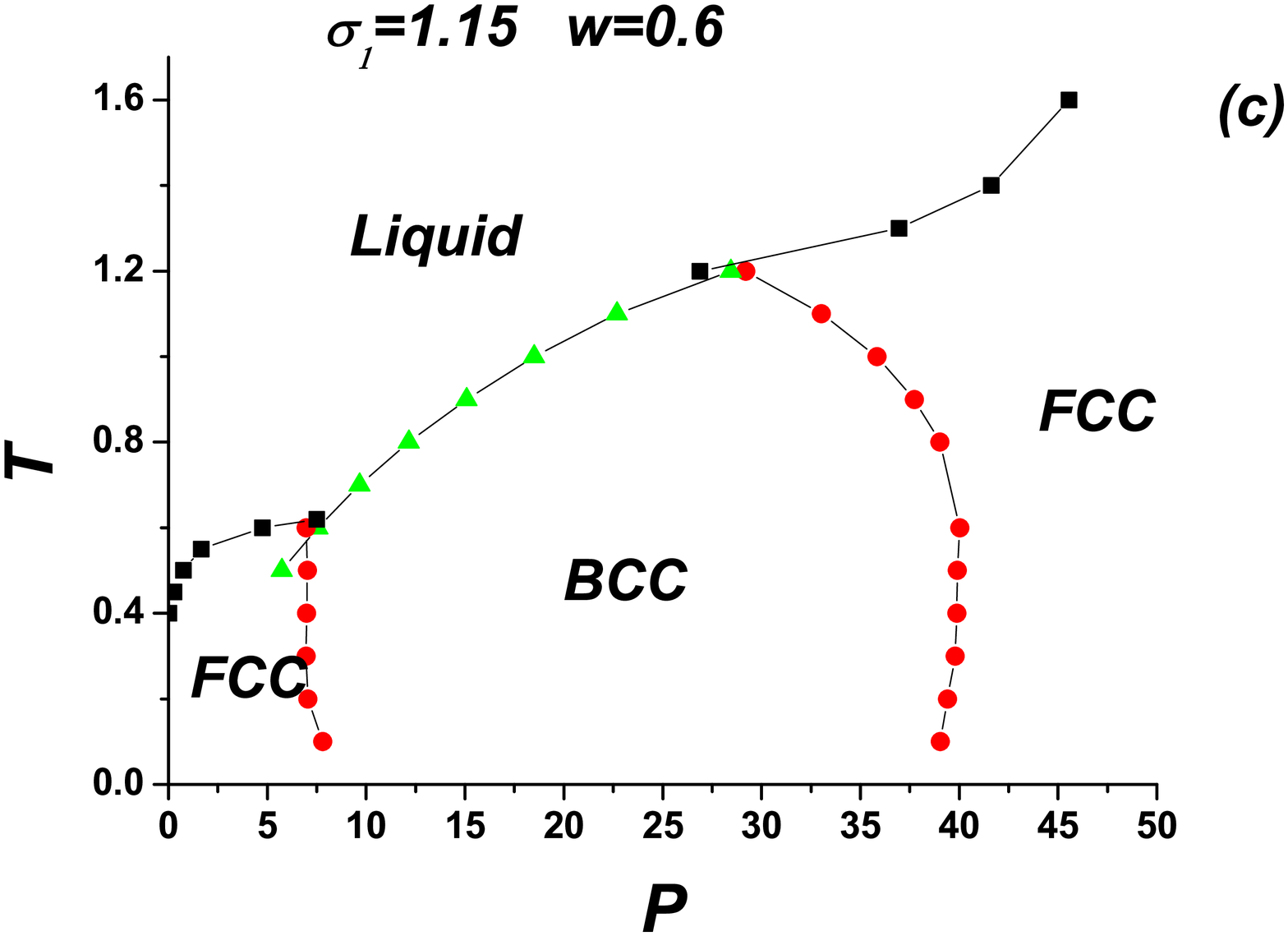}%

\caption{\label{fig:fig4} (Color online) Phase diagrams of the
systems with $\sigma_1=1.15$ (system 1 - 3) in $P - T$
coordinates. a - system 1, b- system 2, c - system 3.}
\end{figure}

In this section we consider the phase diagrams of the systems with
step width equal to $\sigma_1=1.15$ and different depthes of the
attractive well (systems $1-3$ from Table 1). These phase diagrams
are shown in Figs. ~\ref{fig:fig3} (a) - (c) and ~\ref{fig:fig4}
(a) - (c). In our previous publication \cite{wejcp} we reported
the phase diagrams for the purely repulsive potential (zero well
depth). Figs. ~\ref{fig:fig3}(a) and ~\ref{fig:fig4}(a) show these
phase diagrams for the sake of comparison. As one can see from
these figures, the phase diagram consists from liquid and three
crystalline regions: low density FCC (ld-FCC), BCC and high
density FCC (hd-FCC). Note that the shoulder $\sigma_1=1.15$ gives
relatively small perturbation to the soft spheres potential
$\Phi_0(r)=\varepsilon (\frac{\sigma}{r})^{n}$ with $n=14$. As it
was established in \cite{softspfreez} soft spheres freeze into BCC
crystal if $n<6$ and to FCC solid if $n>6$. It means that the
shoulder makes the potential effectively more "soft".

Figs. ~\ref{fig:fig3}(b) and ~\ref{fig:fig4}(b) show the phase
diagrams for the system 2, i.e. the system with $\sigma_1=1.15$
and the well depth $w=0.4$. As it follows from the figures, the
phase diagram is very similar. It again contains liquid region and
three crystalline phases: ld-FCC, BCC and hd-FCC. One can see that
the crystalline region shifts to the higher temperatures comparing
to the previous case. This is clearly connected to the attractive
forces which make the crystal phases more stable.

The phase diagram for the system 3 is shown in Figs.
~\ref{fig:fig3}(c) and ~\ref{fig:fig4}(c). Here the attractive
well is even deeper. It leads to further increase of stability of
the crystalline phases. The melting line shifts to even higher
temperatures. However, the general view of the phase diagram
remains the same.

\begin{figure}
\includegraphics[width=8cm, height=8cm]{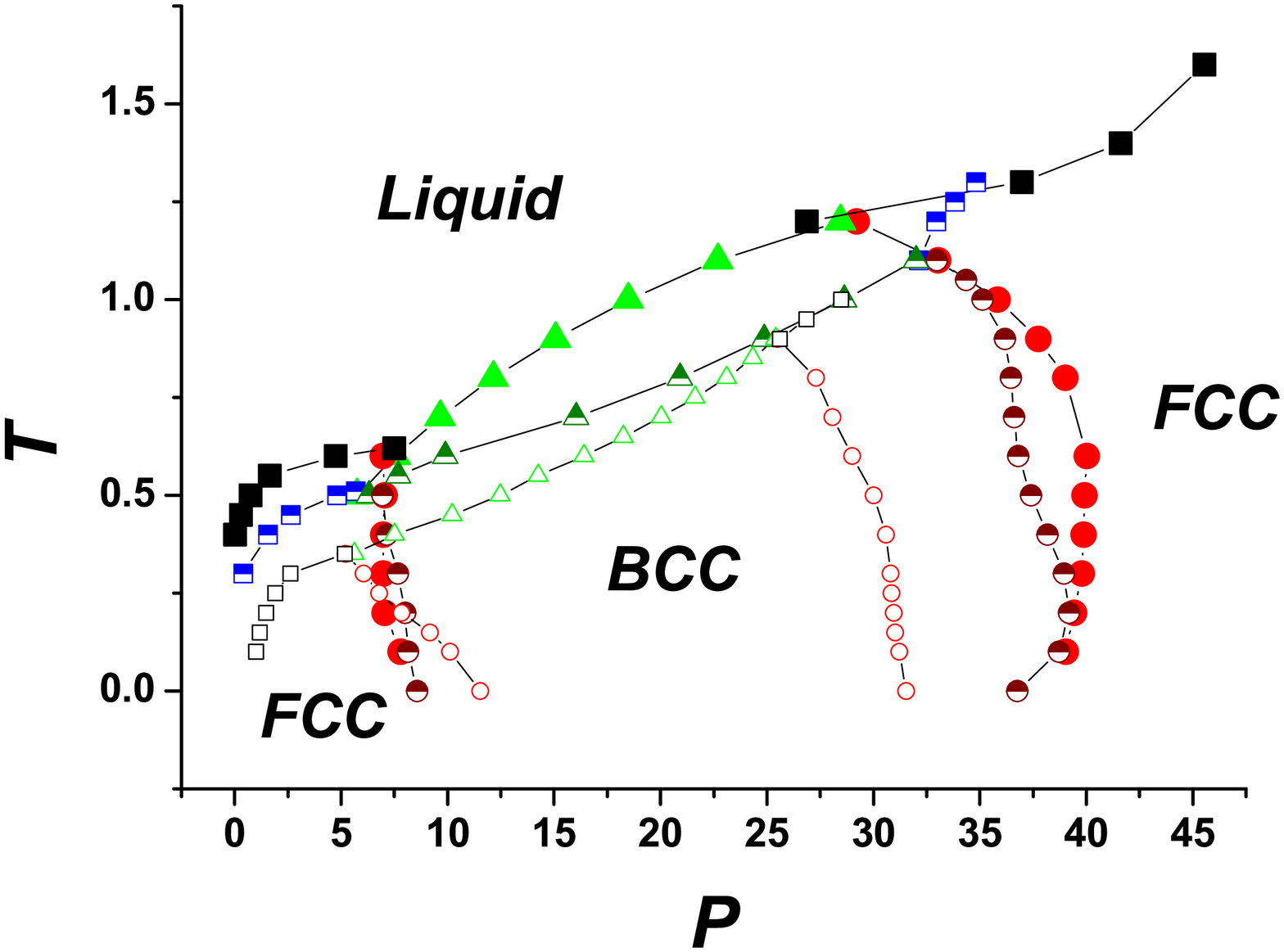}%

 \caption{\label{fig:fig5} (Color online) Phase diagrams of
the systems 1 - 3 in $P - T$ coordinates. Empty symbols - system
1, semifilled symbols - system 2 and full symbols - system 3.}
\end{figure}

Fig. ~\ref{fig:fig5} demonstrates all three phase diagrams in the
same plot for the sake of comparison. Only $P-T$ phase diagrams
are shown since $\rho-T$ plot would be overfilled by the data
points. We observe several trends in the phase diagram evolution
with increasing of the well depth. First of all, in the systems
with attraction the ld-FCC phase is stable already at zero
pressure while in the purely repulsive system positive pressure is
necessary to stabilize the crystal. Secondly the stability of the
crystalline regions with respect to melting grows with increasing
the attractive well: melting temperature at the same density (or
pressure) becomes higher as the attractive well becomes deeper.

The main conclusion of this section is that in the case of the
potentials $1-3$ the attractive well shifts the phase diagram, but
it does not change it's general shape.

\subsection{$\sigma_1=1.35$ (potentials 4 - 7)}

\begin{figure}
\includegraphics[width=8cm, height=8cm]{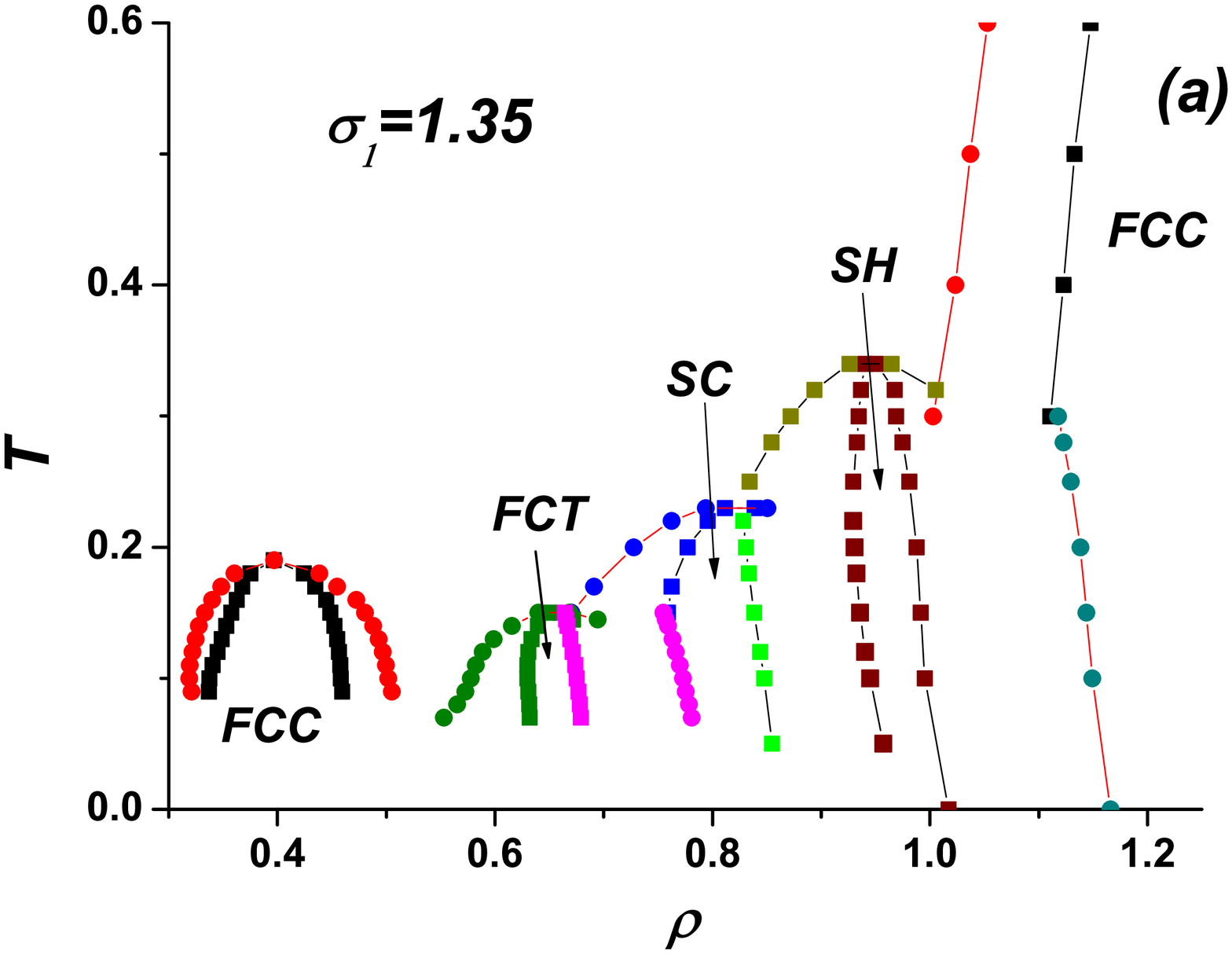}%
\bigskip

\includegraphics[width=8cm, height=8cm]{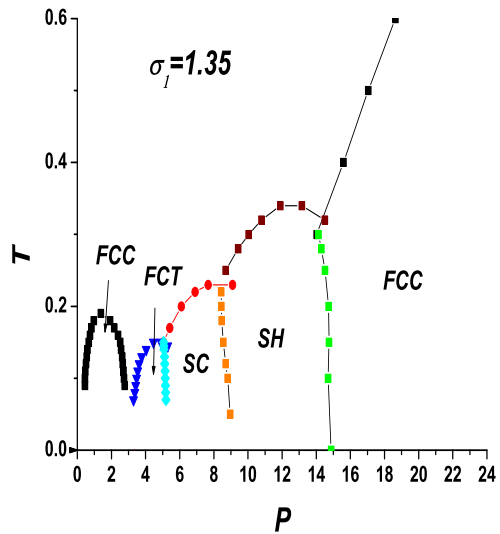}%

\caption{\label{fig:fig6} (Color online) Phase diagram of the
system 4 (Table 1) in (a) $\rho - T$ and (b) $P - T$ coordinates.}
\end{figure}

Now we turn to the potentials with larger width of the repulsive
step $\sigma_1=1.35$. The phase diagram for the purely repulsive
potential with $\sigma_1=1.35$ (potential 4) was published in our
previous paper \cite{wejcp}. After that the zero temperature phase
diagram of this system was reconsidered in the work
\cite{fominpot}. Comparison of our phase diagram (Ref.
\cite{wejcp}) and the zero temperature diagram from Ref.
\cite{fominpot} shows some differences. Body-centered orthorombic
(BCO), body-centered tetrahedral (BCT) and $\beta -Sn$ structures
were reported to be stable into the region where we find
face-centered tetrahedral (FCT) and simple cubic (SC) structures
both at zero and finite temperature. Accordingly to our results
$FCT$ and $SC$ phases are more stable then $BCO$ and $BCT$ in this
region. $\beta -Sn$ was not considered in our work. At the same
time basing on the Ref. \cite{fominpot} we find that simple
hexagonal ($SH$) structure is stable in a certain range of
densities. We carry out the free energy calculations for $SH$
crystal and add the $SH$ transition lines into the phase diagram
(Figs. ~\ref{fig:fig6} (a) - (b)).

The shape of this phase diagram can be intuitively explained as
follows. Since the potential is purely repulsive the overlap of
both hard and soft cores is energetically unfavorable. It leads to
the occurrence of the close packed structure at low densities
(ld-FCC). However, when the pressure (or equivalently density)
increases the particles penetrate the soft core. But since this
core is repulsive, this penetration is energetically unfavorable
and therefore the system forms a set of structures with low
coordination number - FCT (4 nearest neighbors), SC (6 nearest
neighbors) and SH (8 nearest neighbors). Finally, the high density
FCC (hd-FCC) crystal becomes stable which corresponds to the soft
sphere limit.

Taking this into account, one can expect that the ld-FCC phase
should become more stable if the potential has also an attractive
part in addition to the repulsive soft core. One can also expect
that ld-FCC phase occurs al lower pressures comparable to the
purely repulsive potential which corresponds to the arrangement of
particles in the attractive well.

\begin{figure}
\includegraphics[width=8cm, height=8cm]{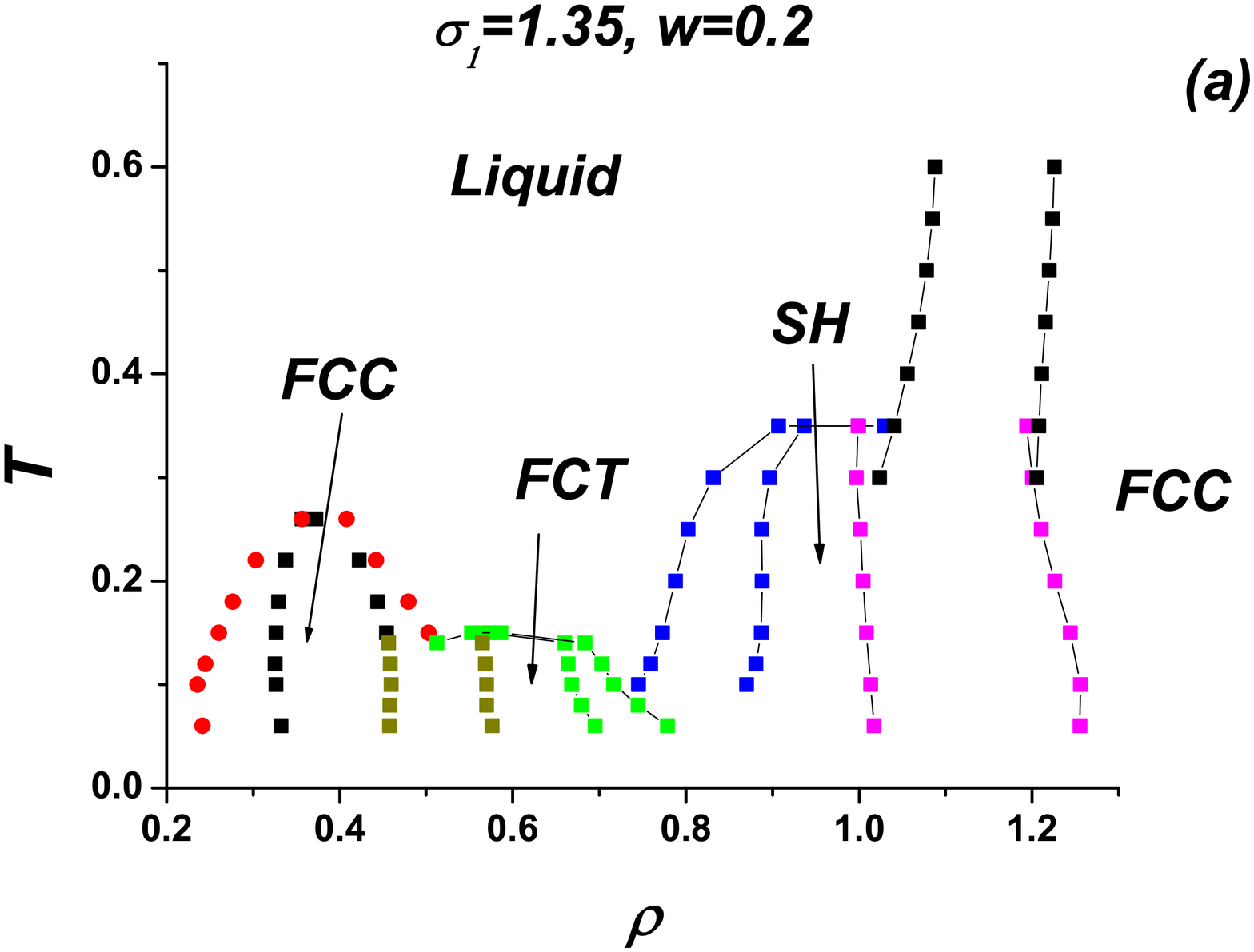}%

\includegraphics[width=8cm, height=8cm]{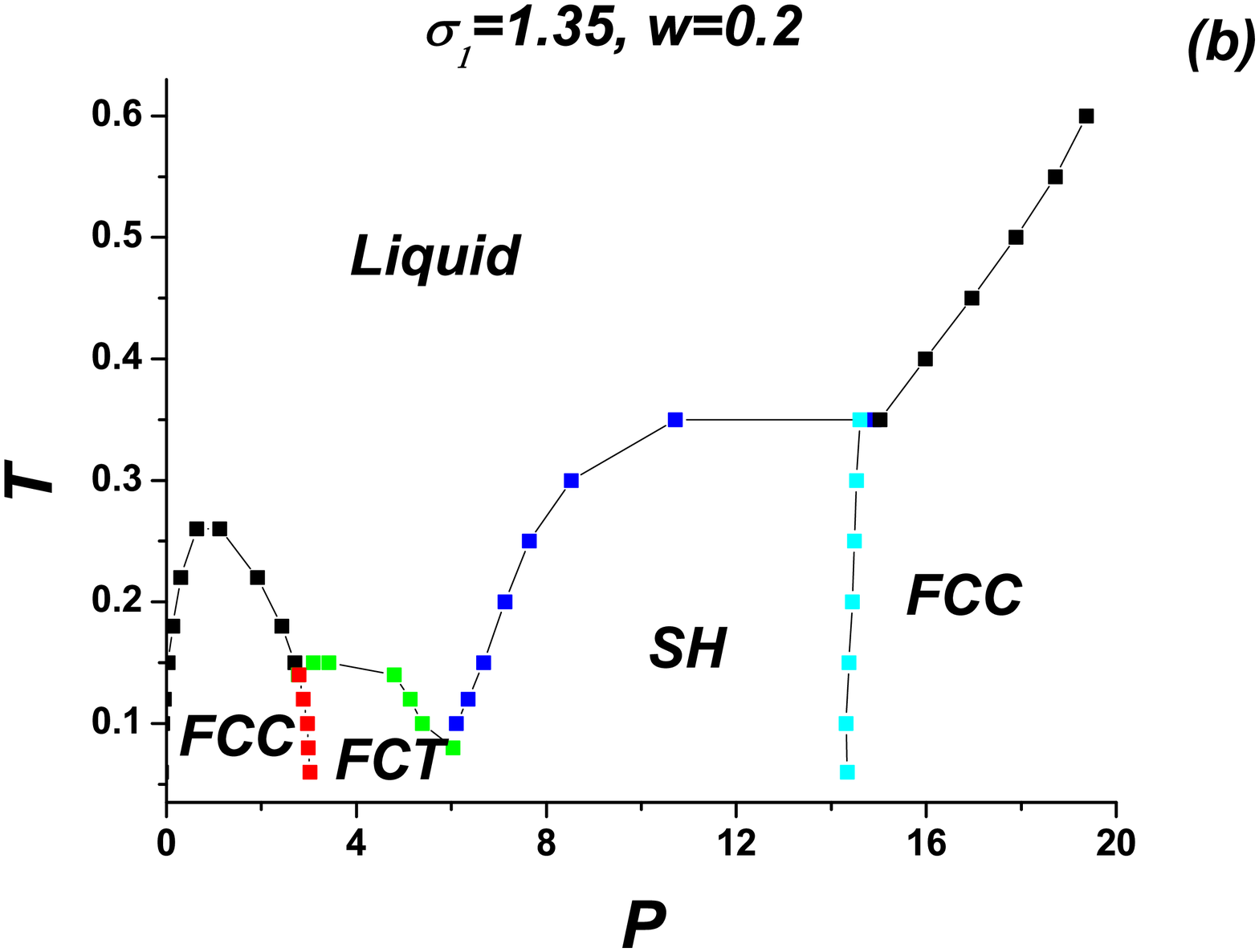}%

\caption{\label{fig:fig7} (Color online) Phase diagram of the
system 5 (Table 1) in (a) $\rho - T$ and (b) $P - T$ coordinates.}
\end{figure}

Figs. ~\ref{fig:fig7} (a) - (b) show the phase diagram for the
system 5, which corresponds to the well depth $w=0.2$.

Comparing the Figs. ~\ref{fig:fig5} and  ~\ref{fig:fig6}, one can
see that the ld-FCC phase do become more stable. While for the
purely repulsive system the maximum temperature of ld-FCC phase is
$T_{max,4}=0.19$, for the system with attraction (system 5) the
maximum temperature is $T_{max,5}=0.27$. At the same time the
region of stability of ld-FCC phase extends and goes into the
negative pressure region.

The FCT phase becomes stable in a wider density range. As a result
the disordered gap presented in the system 4 phase diagram (Figs.
~\ref{fig:fig6} (a) - (b)) is filled by the FCT crystal.

Another change in the phase diagram when the attraction is added
is that the SC crystal disappears. The region of SC crystal is now
filled by FCT phase. The region of the SH phase is almost the same
as in the case of small attraction.

In order to trace the influence of the attraction on the phase
diagram we consider the systems with deeper wells - $w=0.3$
(system 6) and $w=0.4$ (system 7). The phase diagrams of the
system 6 are presented in the Figs. 8(a) and 8(b).

\begin{figure}
\includegraphics[width=8cm, height=8cm]{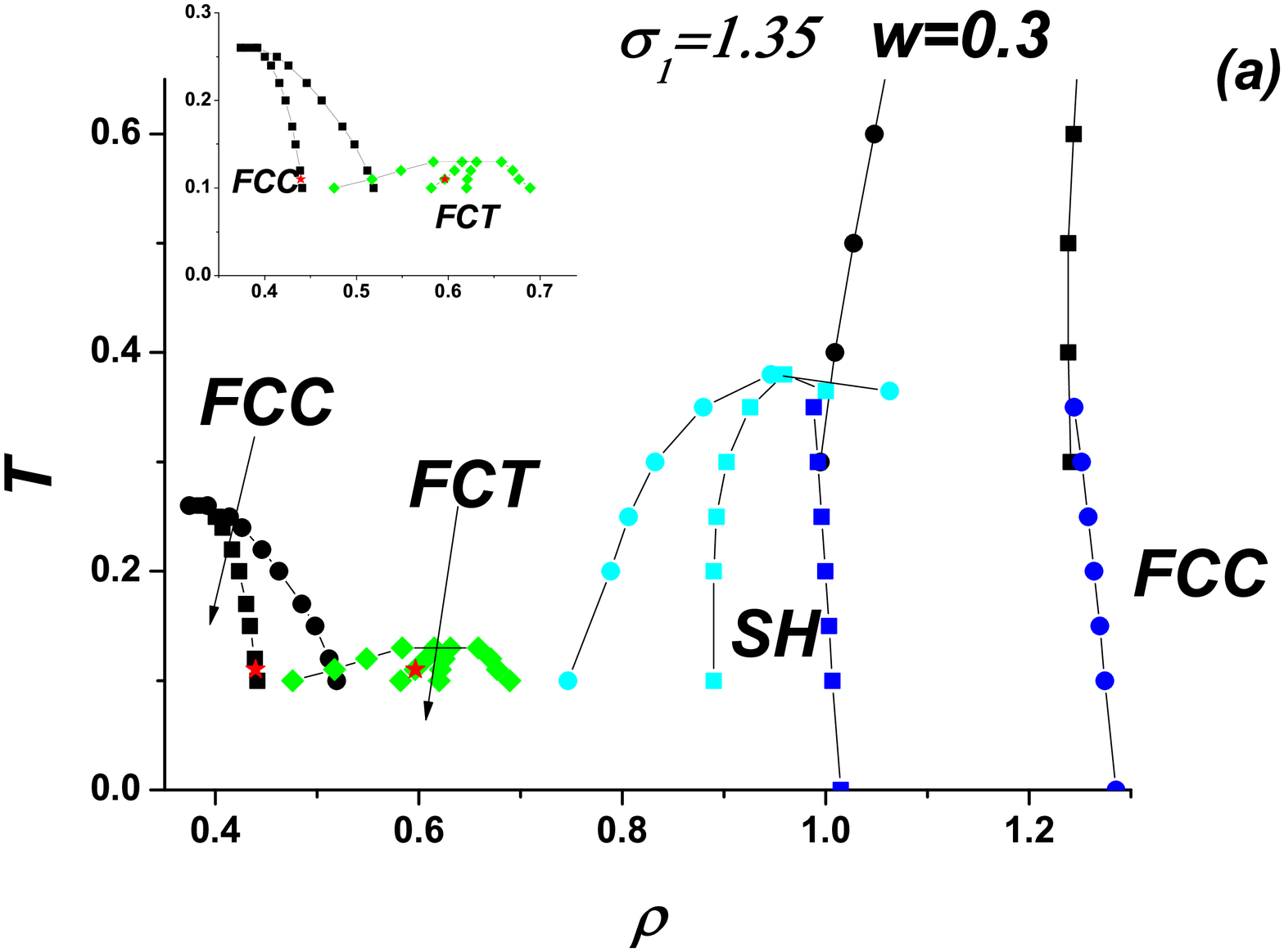}%

\includegraphics[width=8cm, height=8cm]{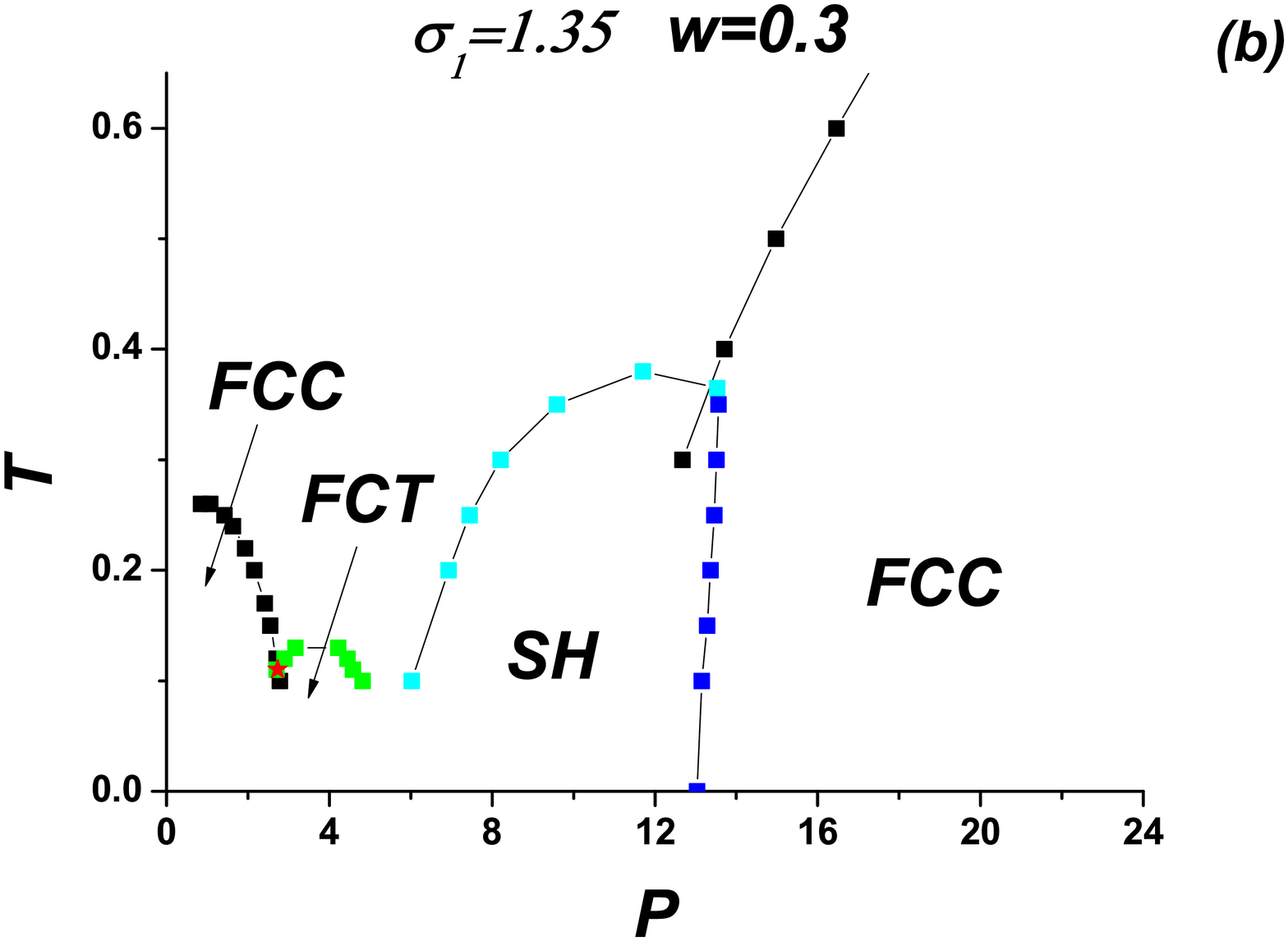}%

\caption{\label{fig:fig8} (Color online) Phase diagram of the
system 6 (Table 1) in (a) $\rho - T$ and (b) $P - T$ coordinates.
The inset of Fig. a shows the l-FCC and FCT phases in large
scale.}
\end{figure}

One can see from Figs. ~\ref{fig:fig8} (a) - (b) that the FCT
phase goes to the lower temperatures region. Also a gap between
FCT and SH crystals appears again. One observes the triple point
ld-FCC - FCT - Liquid at $T=0.11$, but at lower temperatures the
transition ld-FCC - FCT disappears. One can guess that FCT is not
the most thermodynamically stable phase for this system. It means
that if one finds a more stable phase all these contradictions
will be solved. It gives the reasons to suggest that the larger
attraction makes the system to form more complex crystal phases.
The identification of these phases can be a very difficult task
and goes beyond the scope of the present publication.

The ld-FCC phase almost does not change. The maximum on the
melting line stays the same as in the previous case (potential 6).

The range of pressures where SH structure is stable almost does
not change again. This structure shows very weak dependence on the
attraction.

\begin{figure}
\includegraphics[width=8cm, height=8cm]{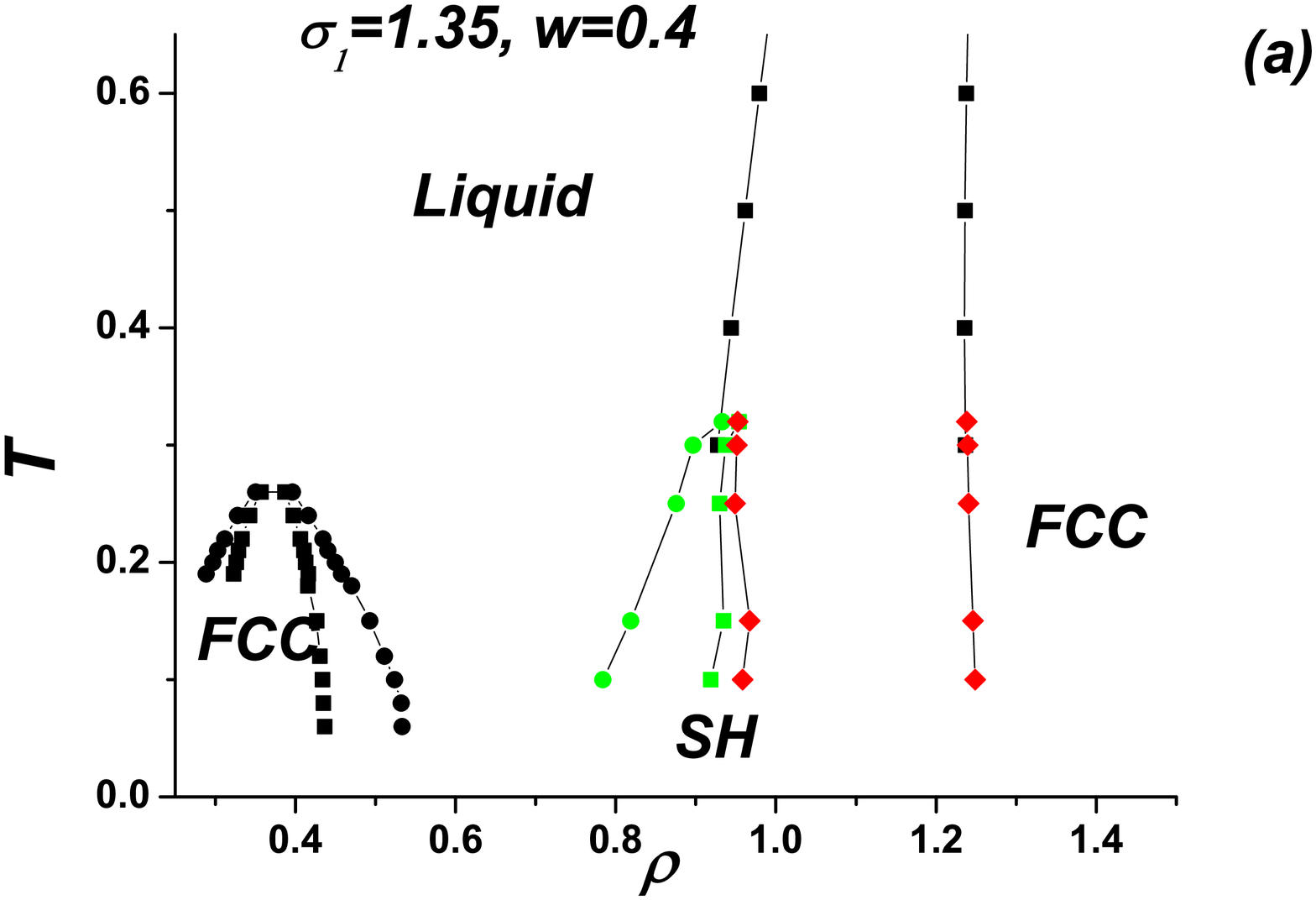}%

\includegraphics[width=8cm, height=8cm]{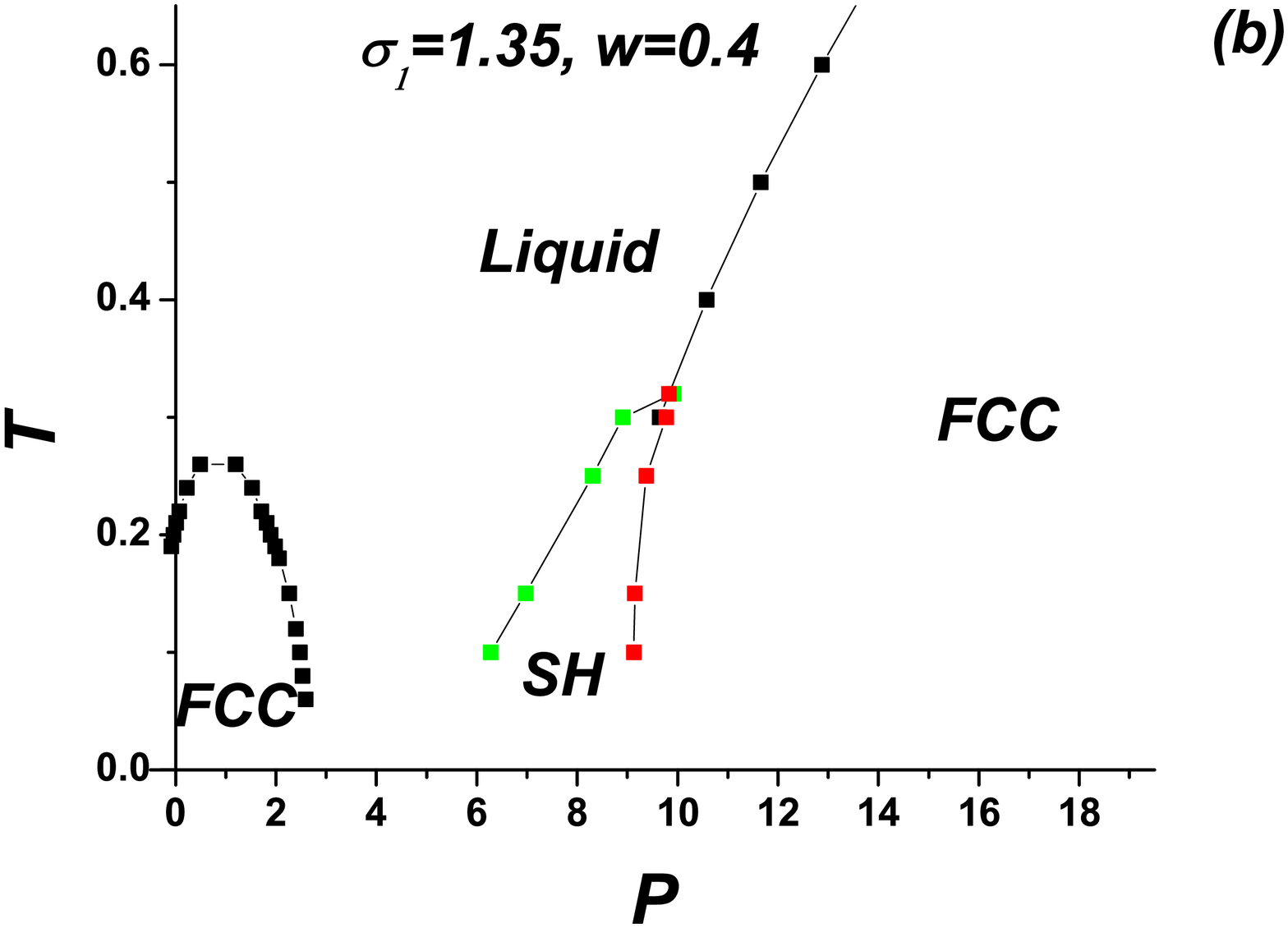}%

\caption{\label{fig:fig9} (Color online) Phase diagram of the
system 7 (Table 1) in (a) $\rho - T$ and (b) $P - T$ coordinates.}
\end{figure}

Figs. ~\ref{fig:fig9} (a) - (b) show the phase diagram of the
system with well depth $w=0.4$ (system 7). One can see from the
figures that FCT phase disappears for this potential. The pressure
range of the ld-FCC and SH structures stay almost unchanged. At
the same time the hd-FCC phase moves to the lower pressures region
which makes the region of SH phase stability more narrow.
Interestingly the SH melting line looses the maximum observed for
the previous systems. One can guess that this maximum goes under
the hd-FCC melting line and becomes metastable. As a result, since
SH phase does not reach its maximum, the temperature range of
stability of this structure slightly decreases - from $T_{max,6}
^{sh}=0.38$ to $T_{max,7} ^{sh}=0.32$. One can expect that, if the
attraction becomes even stronger, the SH phase goes into the
hd-FCC region and becomes metastable with respect to it. A large
gap between ld-FCC and SH phases occurs. Most probably this gap is
filled by some unknown structure as it was discussed above for the
system 6.

\begin{figure}
\includegraphics[width=8cm, height=8cm]{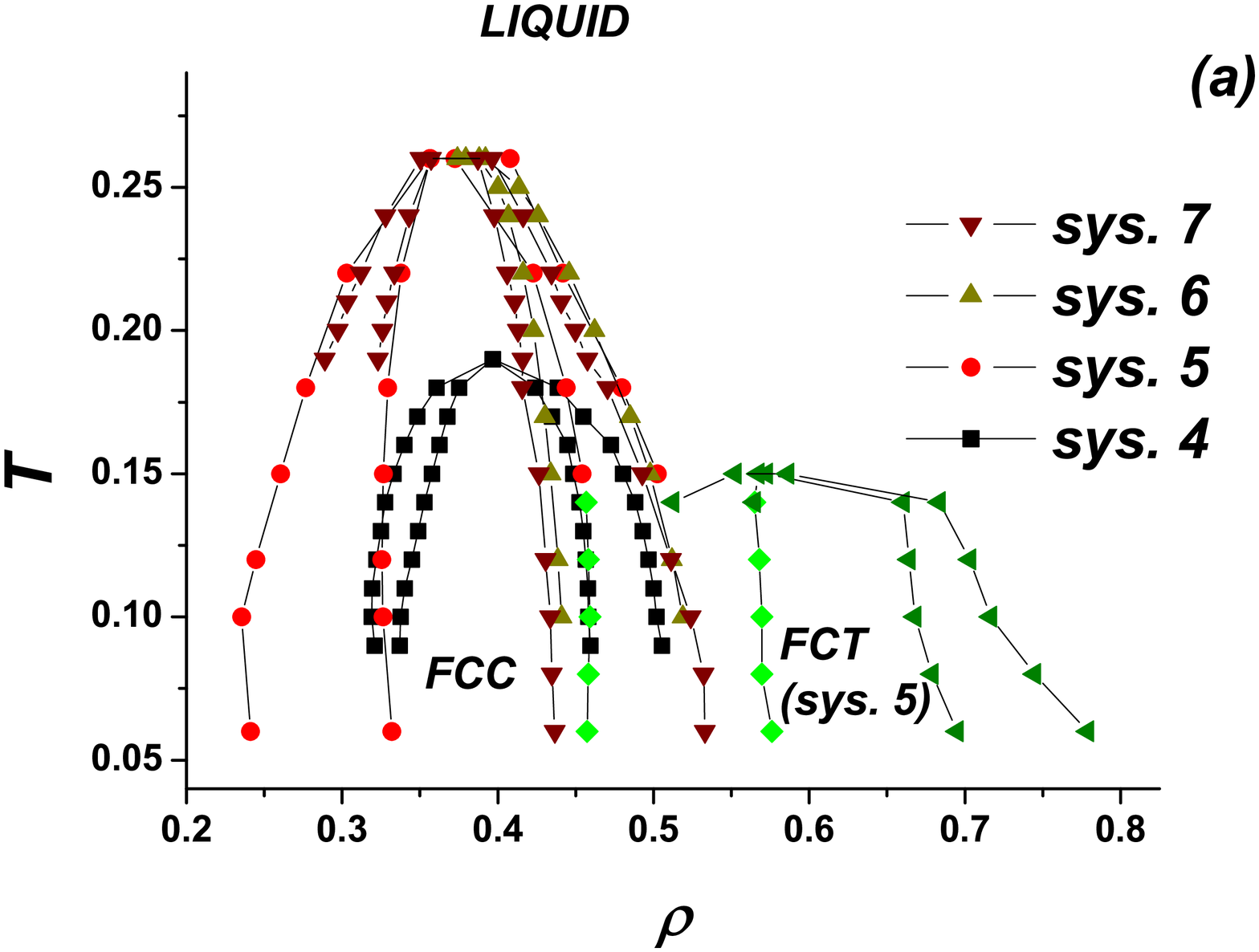}%

\includegraphics[width=8cm, height=8cm]{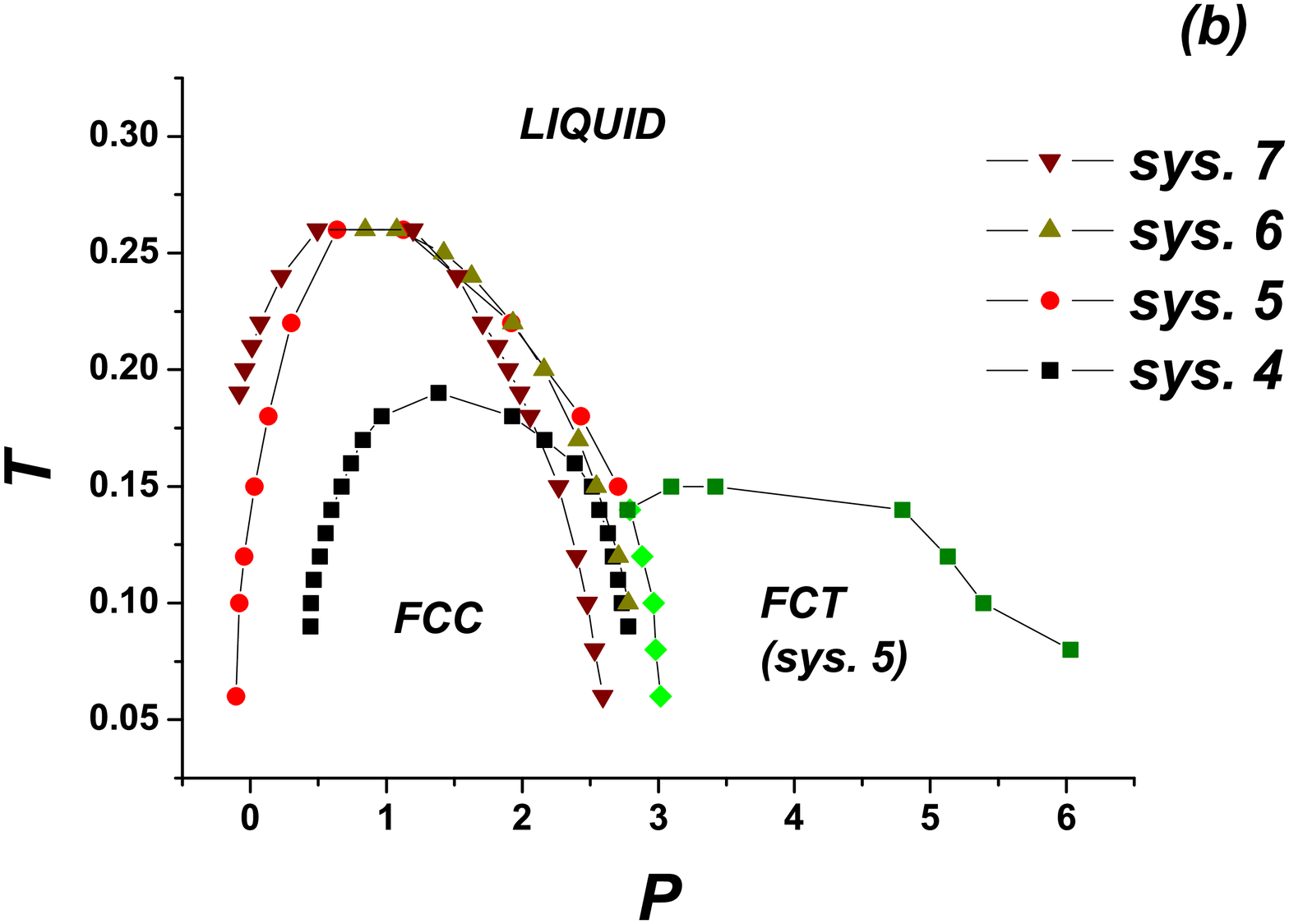}%

\caption{\label{fig:fig10} (Color online) ld-FCC phases for the
systems 4-7 (Table 1) in (a) $\rho - T$ and (b) $P - T$
coordinates.}
\end{figure}

It is worth to trace the evolution of the ld- and hd-FCC phases
with increasing attraction. From Figs. ~\ref{fig:fig10} (a) - (b)
one can see that when the attraction is added the ld-FCC phase
becomes more stable with respect to temperature. However,
increasing of the attraction does not stabilize the ld-FCC phase
any more. At the same time the attraction moves the region of
ld-FCC stability to the negative pressures. One can imagine that
with further increase of attractive part the melting line will
have negative slope even at zero pressure which qualitatively
corresponds to the case of water.

\begin{figure}
\includegraphics[width=8cm, height=8cm]{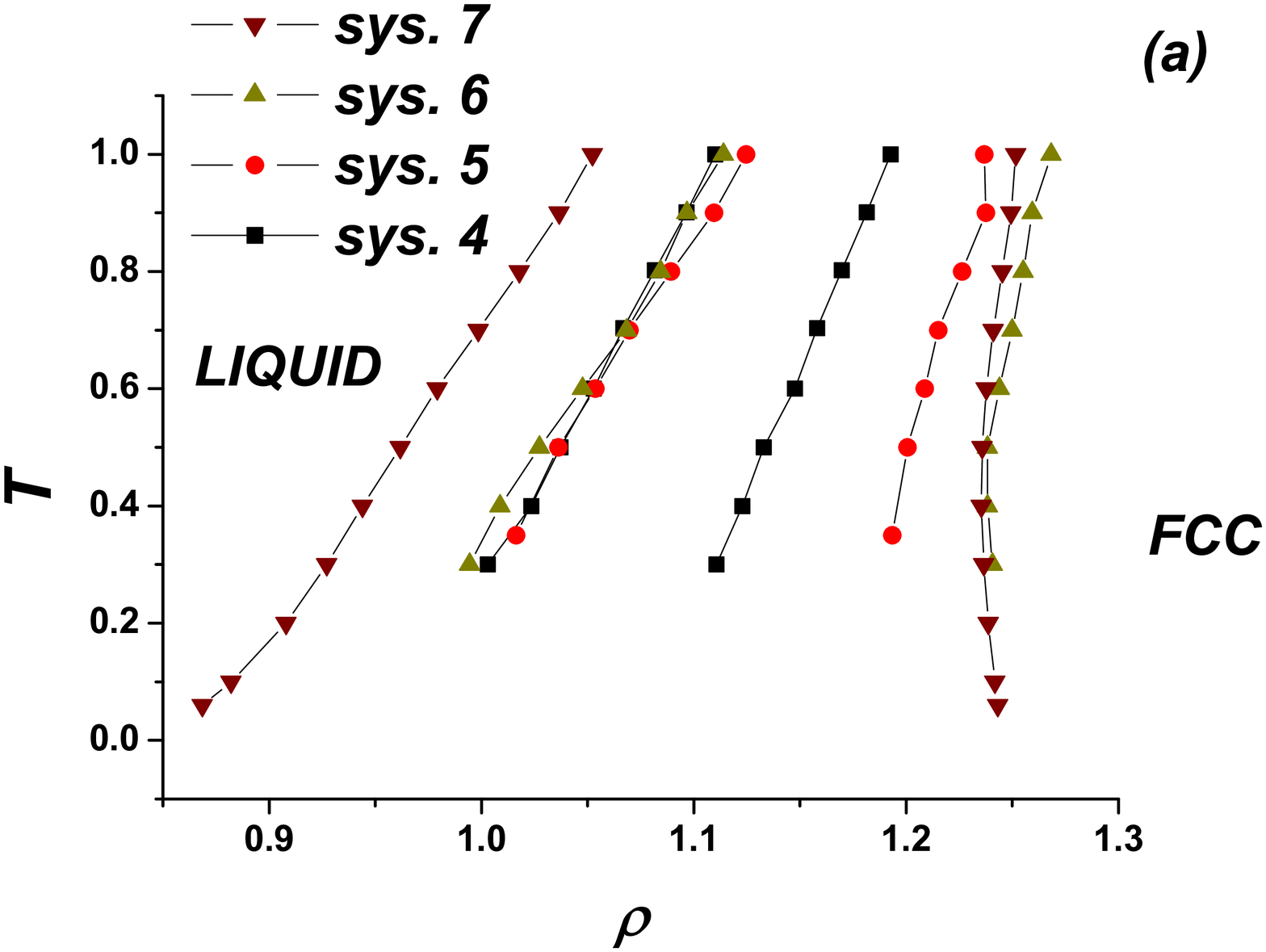}%

\includegraphics[width=8cm, height=8cm]{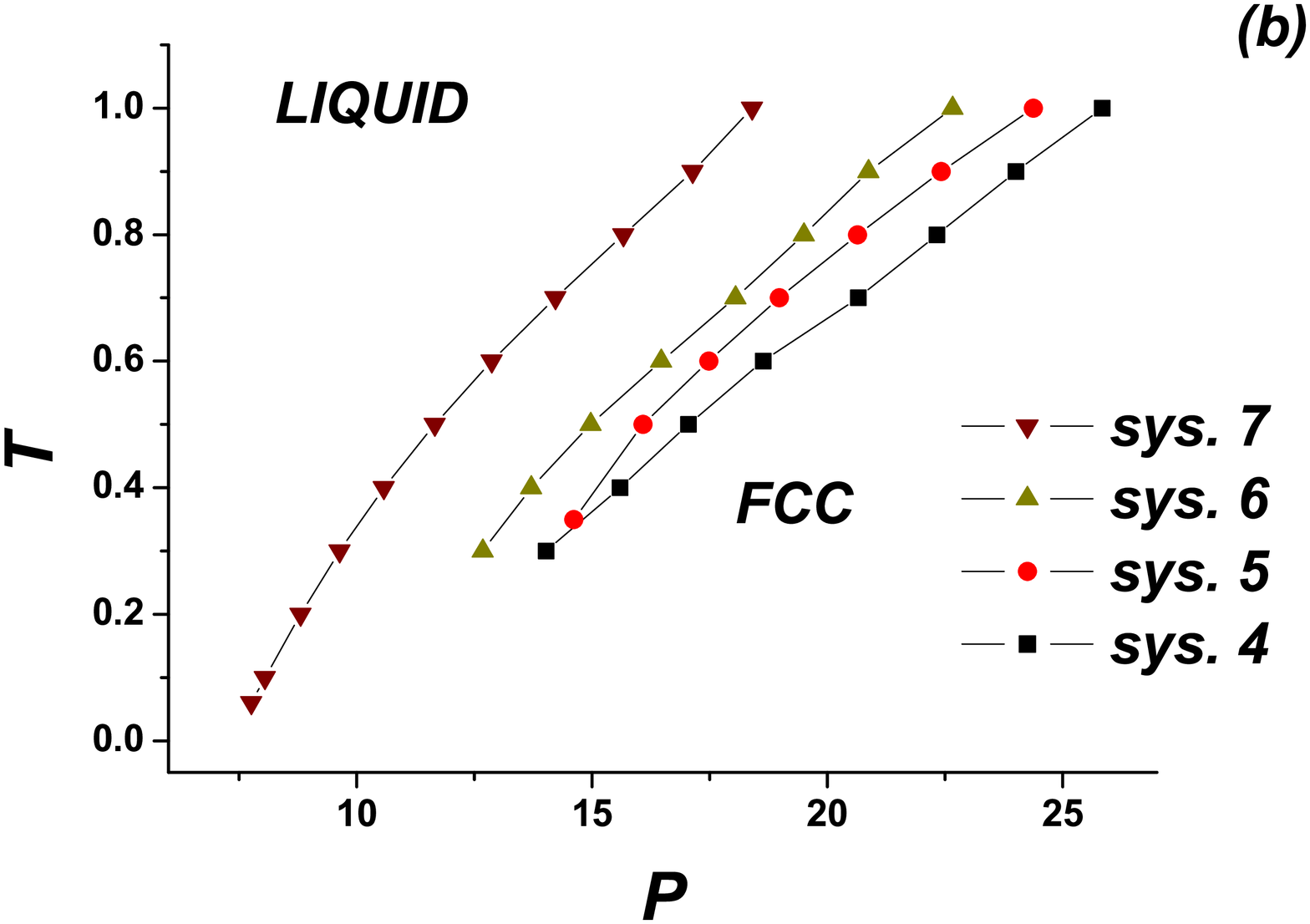}%

\caption{\label{fig:fig11} (Color online) hd-FCC phases for the
systems 4-7 (Table 1) in (a) $\rho - T$ and (b) $P - T$
coordinates.}
\end{figure}

Attraction has much stronger effect on the hd-FCC melting line
(Fig. ~\ref{fig:fig11} (a) - (b)). Small attraction ($w=0.2$,
system 5) almost does not affect the hd-FCC melting line, while
increasing of the attraction moves it to the lower pressures. It
makes the gap between ld-FCC and hd-FCC narrower. One can expect
that at some value of the well depth the gap disappears and the
crystal region of the phase diagram consists from two FCC phases
only.

\begin{figure}
\includegraphics[width=8cm, height=8cm]{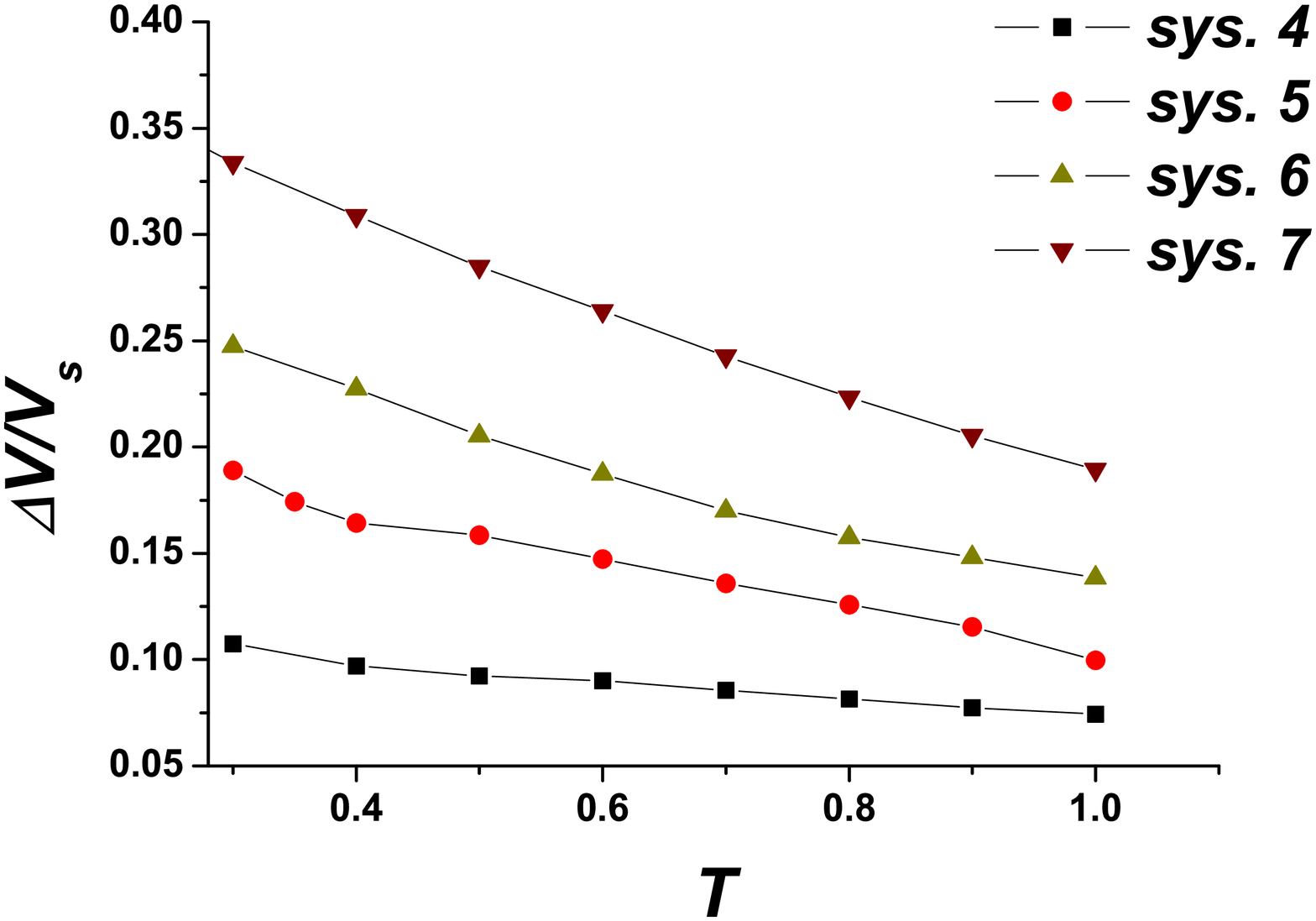}%

\caption{\label{fig:fig12} (Color online) Volume jump over the
crystal state specific volume for hd-FCC phase of the system
$4-7$}
\end{figure}

Fig. ~\ref{fig:fig12} shows the specific volume jump over the
crystal phase volume at hd-FCC melting for the systems $4-7$.
Interestingly, the volume jump increases considerably with
increasing the attractive well and reaches the value up to
$\approx 33 \%$. Experimental volume jumps are usually of the
order of $5-10 \%$, although for some substances it can reach much
higher values.

The example of the volume jump clearly shows the importance of the
parameter $\varepsilon_2 /\varepsilon_1$, where $\varepsilon_1$ is
the step height and $\varepsilon_2$ - well depth. Even at the same
length parameters - $\sigma_1/ \sigma$ and $\sigma_2/ \sigma$ the
character densities in the system can be strongly different.

\section{IV. Thermodynamic Anomalies}

Finally we would like to discuss anomalous behavior of the core
softened systems. In our previous works
\cite{wejcp,wepre,weros,wearXiv} we showed that purely repulsive
system (system 4) demonstrates diffusion and density anomalies.
Here we extend this study to the RSS-AW system.

The problem of anomalous behavior of isotropic core softened
system with attraction has already been discussed by several
authors \cite{franz1,franz2,barban}.

In the recent publications \cite{franz1,franz2} the characteristic
length and energies of the interatomic potential were fixed while
the softness of the potential varied. It means that the focus of
these publications is slightly different comparing with the
present one. The authors found anomalous behavior in the system
they studied in a wide range of the potential parameters.

In the publication \cite{barban} the depth of attractive well is
varied with all other parameters fixed. In this sense this work is
very close to the present one. The deepest well considered in this
work is $-1.0$ which is deeper then in our study. The main
conclusion of the authors is that the anomalies shrink with
increasing the well depth and finally disappear.

\begin{figure}
\includegraphics[width=8cm, height=8cm]{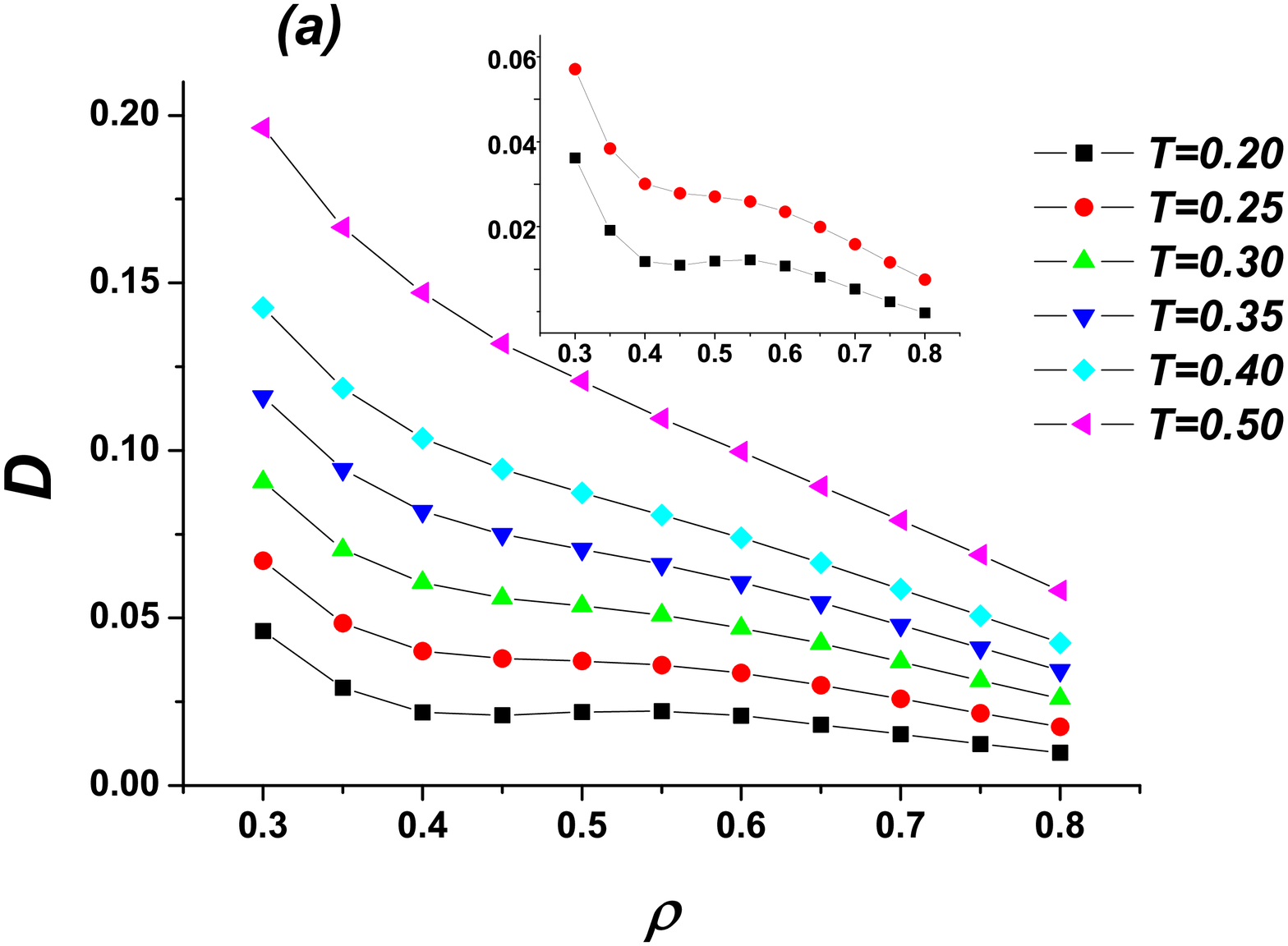}%

\includegraphics[width=8cm, height=8cm]{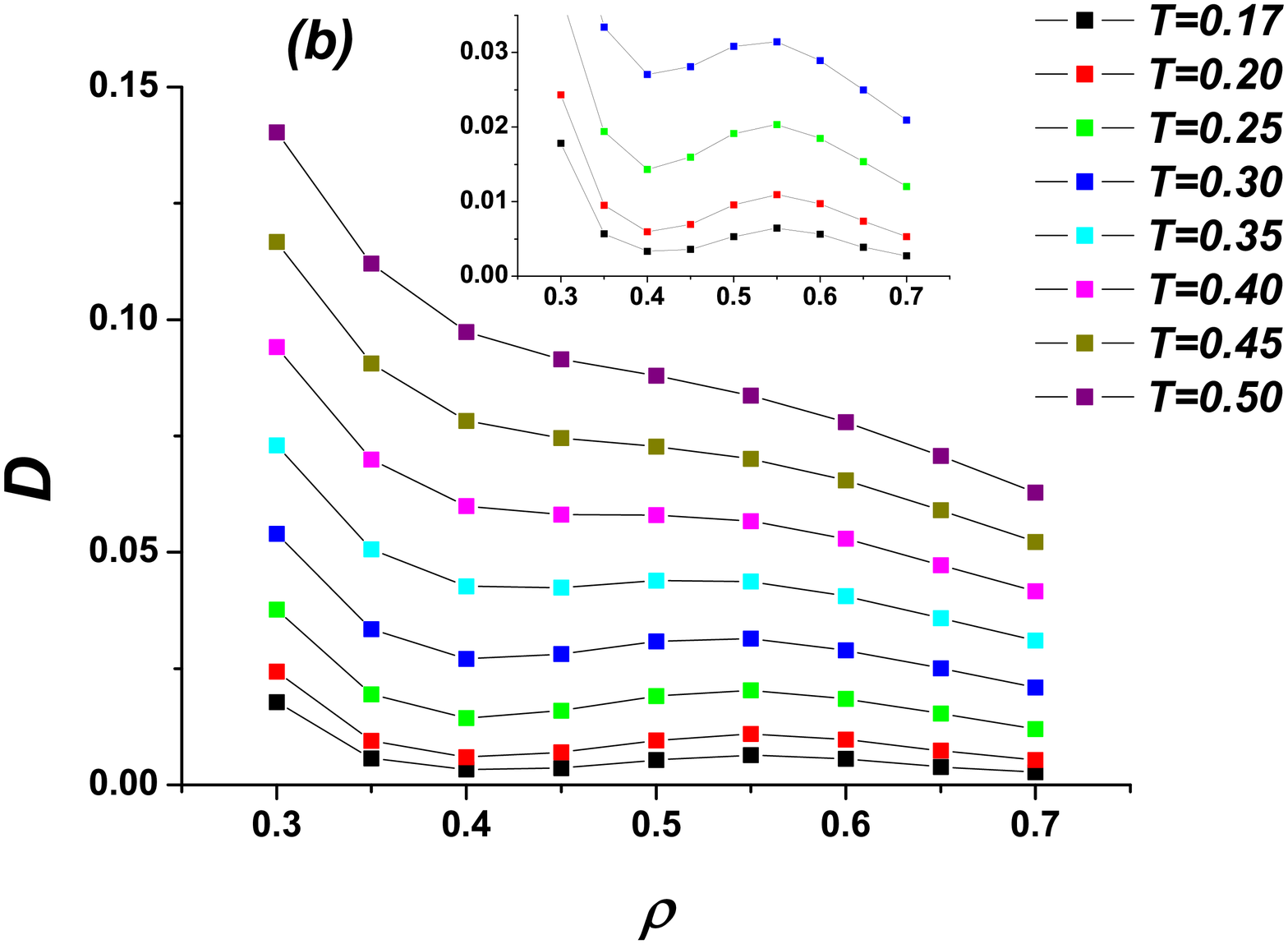}%

\includegraphics[width=8cm, height=8cm]{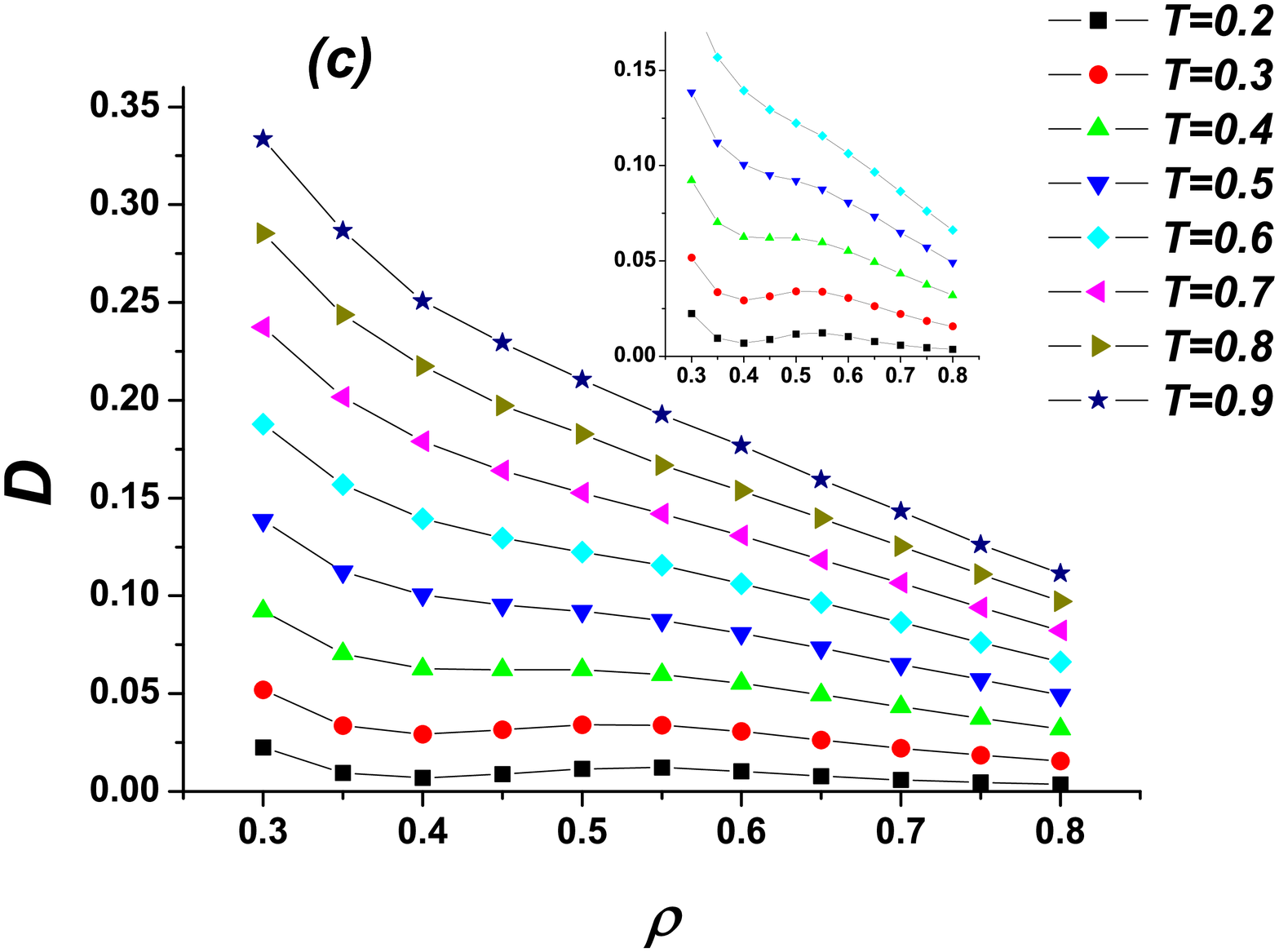}%
\caption{\label{fig:fig13} (Color online) Diffusion coefficient
for (a) purely repulsive system with $\sigma_1=1.35$ (system 4)
and systems with attraction - (b) $\sigma_1=1.35$ and $w=0.2$
(system 5), (c) $\sigma_1=1.35$ and $w=0.4$ (system 7). The insets
show low-temperature regimes in larger scale.}
\end{figure}

Figs.~\ref{fig:fig13} (a)-(c) present the diffusion anomaly for
the systems $4$, $5$ and $7$. One can see from these figures that
the density range of anomalous diffusion does not change on
increasing the well depth. At the same time the maximum
temperature of anomalous behavior increases. It means that small
attraction stabilizes anomalous diffusion in the system.

In order to see the origin of the difference in the qualitative
behavior of diffusion anomaly region between the present work and
the publication \cite{barban} we should notice that in our case
the increasing of the well depth at the same time makes the
potential softer (Fig.~\ref{fig:fig2}). This corresponds to making
the parameter $\Delta$, which controls the slope of the potential
in the works \cite{franz1,franz2}, smaller. Accordingly to these
publications the anomalous region is wider for the smaller values
of this parameter. Seems that this effect is larger in our case
then shrinking of the anomalous region with increasing attraction.

More detailed discussion of the anomalous behavior of the RSS-AW
systems will be given in a subsequent publication.

\section{IV. Conclusions}

The present article shows the evolution of the phase diagrams of
two core-softened systems with addition of attractive forces. It
is shown that the general shape of the phase diagram and the set
of preferable structures is mostly defined by repulsive shoulder
while attractive well just shifts the phase diagram in $\rho - T$
plane. Anomalous diffusion behavior is also considered. It is
shown that the appearance of the diffusion anomaly if related to
the soft-core but the small attraction stabilizes it at higher
temperatures.

\bigskip

\begin{acknowledgments}
We thank V. V. Brazhkin and Daan Frenkel for stimulating
discussions. Y.F. also thanks the Russian Scientific Center
Kurchatov Institute for computational facilities. The work was
supported in part by the Russian Foundation for Basic Research
(Grants No 08-02-00781 and No 10-02-00700) and Russian Federal
Program 02.740.11.5160.
\end{acknowledgments}


\end{document}